\documentclass[%
reprint,
amsmath,amssymb,
prb,
nofootinbib, superscriptaddress]{revtex4-2}
\usepackage{etoolbox}
\makeatletter
\patchcmd{\@footnotetext}
  {\setspace@singlespace}{0.5}
  {}{}
\makeatother
\usepackage{graphicx}% Include figure files
\usepackage{subcaption}
\usepackage{mwe}
\usepackage{dcolumn}% Align table columns on decimal point
\usepackage{bm}% bold math
\usepackage{listings}
\usepackage[utf8]{inputenc}
\usepackage{wrapfig}
\usepackage{hyperref}
\hypersetup{
    colorlinks=true,
    linkcolor=blue,
    filecolor=blue,      
    urlcolor=blue,
    pdftitle={Photonic Quantum Computing For Polymer Classification},
    pdfpagemode=FullScreen,
    citecolor=blue,
    }

\graphicspath{ {./Images_WP2/} }
\usepackage{mathtools}

\DeclarePairedDelimiter\ket{\lvert}{\rangle}
\DeclarePairedDelimiterX\braket[2]{\langle}{\rangle}{#1 \delimsize\vert #2}

\usepackage[mathlines]{lineno}% Enable numbering of text and display math
\usepackage[normalem]{ulem}

\usepackage[dvipsnames]{xcolor}

\definecolor{bpc}{RGB}{0,170,80}

\begin{document}
\title{Photonic Quantum Computing For Polymer Classification}
\date{\today}

\author{Alexandrina Stoyanova}
\thanks{These two authors contributed equally to this work and share first authorship.}
\affiliation{Alysophil, 850 Bd Sébastien Brant, 67400 Illkirch-Graffenstaden, France}
\author{Taha Hammadia}
\thanks{These two authors contributed equally to this work and share first authorship.}
\affiliation{Quandela, 7 Rue Léonard de Vinci, 91300 Massy, France}
\affiliation{École Polytechnique, Rte de Saclay, 91120 Palaiseau, France}
\author{Arno Ricou}
\affiliation{Quandela, 7 Rue Léonard de Vinci, 91300 Massy, France}
\author{Bogdan Penkovsky}
\affiliation{Alysophil, 850 Bd Sébastien Brant, 67400 Illkirch-Graffenstaden, France}

\begin{abstract}
We present a hybrid classical-quantum approach to the binary classification of polymer structures. Two polymer classes visual (VIS) and near-infrared (NIR) are defined based on the size of the polymer gaps. The hybrid approach combines one of the three methods, Gaussian Kernel Method, Quantum-Enhanced Random Kitchen Sinks or Variational Quantum Classifier, implemented by linear quantum photonic circuits (LQPCs), with a classical deep neural network (DNN) feature extractor.~The latter extracts from the classical data information about samples chemical structure. It also reduces the data dimensions yielding compact 2-dimensional data vectors that are then fed to the LQPCs. We adopt the photonic-based data-embedding scheme, proposed by Gan \textit{et al.} [EPJ Quantum Technol. \textbf{9}, 16 (2022)] to embed the classical 2-dimensional data vectors into the higher-dimensional Fock space. This hybrid classical-quantum strategy permits to obtain accurate noisy intermediate-scale quantum-compatible classifiers by leveraging Fock states with only a few photons.~The models obtained using either of the three hybrid methods successfully classified the VIS and NIR polymers.~Their accuracy is comparable as measured by their scores ranging from 0.86 to 0.88. These findings demonstrate that our hybrid approach that uses photonic quantum computing captures chemistry and structure-property correlation patterns in real polymer data.~They also open up perspectives of employing quantum computing to complex chemical structures when a larger number of logical qubits is available.
\end{abstract}

\maketitle

\section{Introduction}\label{Introduction}
Polymers are stochastic mixtures of macromolecules with complex chemical and topological structures and dynamics.~The structures cover a vast range of length scales, from monomer size (\AA) to the radius of gyration (nm) and microstructures ($\mu$m). The dynamics typically covers time scales from femto-seconds to minutes and even years in the case of aging, vitrification, jamming, and semi-crystallization processes \cite{ref_multi_scale, ref1, ref2, data_driven}.~This complexity gives rise to properties like high thermal and electric conductivity, corrosion and high-temperature resistance, elasticity, or biodegradability.~Polymer properties can further be tuned by dispersing inorganic nanoparticles into the polymer matrix in the so-called nanocomposites (see, e.g.,~\cite{ref2}). Both natural and synthetic polymers are ubiquitous in industries like electronics, cosmetics, drug design, carbon capture, aerospace, automobiles, clothing, communication,  organic photovoltaic, and 3D printing.  

Polymer physical, mechanical, and chemical properties are governed by their chemical structure, monomer arrangement, chain size\footnote{A mixture of chains with different lengths is characterized by a molecular weight distribution.}, morphology, and conformations. Inter- and intra-molecular interactions between the chains can also produce micro-structures with different thermal and mechanical properties related, for example, to phase separations (amorphous and crystalline regions in a polymer).~The development of optimal polymerization processes for new polymers and composites that can be time and resource consuming involves, thus, an in-depth understanding of polymer hierarchical structures. Such a knowledge can be acquired from the interplay of material informatics, data science, polymer physics, and chemistry.~Computational approaches are thereby employed to gain insight into the polymer structures and even generate additional data \cite{ref_multi_scale, ref1, ref30, ref33, ref34}.~Exploring the entire polymer chemical space is, however, impossible for any existing model or computational method on its own.~Instead, only parts of that space are explored while searching for macromolecules with specific properties. 

~Existing models and methods (see, e.g., Review \cite{ref1}) operate each at different length and time scales. %since the separation of those scales is inherent to polymers. 
For example, the structure and dynamics at length scales of the order of a chain size might be unaffected by the monomer's chemistry (atomic types and bonding).~Polymer lattice models can then be adopted, modeling the polymer by a chain of connected beads \cite{ref1, ref3, LM1}\footnote{A bead or a grain is an assemble of atoms, fragments or entire monomers that is treated as a whole in a calculation for the polymer structure.~The bead is a building block in coarse-grained models for polymers.} placed on a lattice to represent chain conformations. Little or no information about the monomer's chemistry is thereby used.~If the studied properties are, on the other hand, intrinsic to the monomer's chemistry like the band gap, electronic polarizability, and dielectric constant, the atomic ($\sim$10$^{-9}$ m, $\sim$10$^{-9}$--10$^{-6}$ s) and even quantum scales ($\sim$10$^{-10}$ m, $\sim$10$^{-12}$ s) are implicated.~A quantum treatment is then the most appropriate one, since the macromolecules are quantum-mechanical objects, but without important model simplifications   \cite{QC_polymers} the computational costs can be well beyond current resources, even when the latter are enabled by Graphics Processing Units (GPUs). Furthermore, phenomena at different scales need often be treated at equal footing while describing polymer properties.    
 Hence, efforts have been invested into developing multi-scale models and simulations \cite{ref_multi_scale, ref1, ref23, ref31, ref32}.~A major challenge for such strategies is finding proper protocols to combine methods that operate at different length and time scales. 
 
Data-driven Machine Learning (ML) or Deep Learning (DL) algorithms have also been leveraged recently for property prediction and $in~silico$ synthesis of new polymers \cite{data_driven, ref4, ref5, ref6, ref7, ref8, ref9, ref10, ref11, ref12, ref21, ref22, ref23, ref25, ref28, ref30}.~Key ingredients are property- and structure-related data and machine-readable polymer features.~Experimental synthetic polymer data bases are, however, sparse and heterogeneous, as measuring the polymer properties is resource and time consuming.~Furthermore, converting the hierarchical polymer structures into chemically-informed representations and then into machine-readable features is not straightforward.~A number of well-defined representations such as SMILES\footnote{Simplified Molecular Input Line Entry Specification}, graphs, binary images, molecular descriptors, and fingerprints exist for molecules with deterministic structures (see, Review \cite{ref35}). However, only a few approaches \cite{ref9, ref13, ref21} have been conceptualized for featurizing the stochastic polymer structures.~Depending on the studied property, a featurization scheme might only operate at a certain length scale. For example, predicting the dielectric constant requires an adequate numerical representation of the monomer's chemical structure.~On the other hand, the featurization scheme can ignore the chains morphology and conformation because of their minor impact on the dielectric constant. For other properties like the temperature of glass transition that is expected to depend on the polymer conformations \cite{ref2}, more complex featurizations are needed.     

The intricate structure-property correlation patterns in polymer data that need to be reflected by complex featurization schemes can be harder to detect by classical machines running classical ML or DL learning algorithms, even when the former
are empowered by GPUs. Quantum computing on quantum hardware, on the other hand, allows calculations in higher-dimensional spaces and thus enables the use of more complex learning models and numerical representations. Polymer materials are therefore an ideal playground for quantum algorithms because of their large chemical space (see, e.g., \cite{Large_CS_polymers,Large_CS_polymers_2}).

In this work, we explore the performance for polymer classification tasks of three quantum machine learning models: Gaussian Kernel Method (GKM), Random Kitchen Sinks (RKS), and Variational Quantum Classifier (VQC) implemented  using linear Quantum Photonic Circuits (QPCs) \cite{ref42}. We introduce a hybrid classical-quantum approach that combines a classical Deep Neural Network (DNN) feature extractor with each of the three formalisms. The classical DNN feature extractor extracts chemically-relevant information about the polymer species. It also reduces the dimensionality of the classical data which is then embedded into the higher-dimensional Fock space.~In fact, the numerical representations used for chemical species often utilize high-dimensional vectors of either integers or floating-points numbers. Encoding those directly on qubits would require an important number of logical qubits and photons, respectively. The photonic-based   
Quantum Machine Learning (QML) models exploited here circumvent this constraint by mapping the classical data points into the high-dimensional Fock space\footnote{The dimension of the Fock space grows exponentially with the number of photons $n$ and spatial modes $m$.}. This strategy allows for using simpler input Fock states with two-mode linear QPCs and a small number of photons and still building accurate quantum classifiers. The additional dimensionality reduction operated by the classical DNN feature extractor facilitates the strategy. In fact, in the NISQ era of a small to intermediate number of qubits, a classical pre-processing of classical high-dimensional data is a necessary step that can involve compact features extraction \cite{maria, TransferLearning, Quantum_Preskill}.      Furthermore, reducing the number of photons \footnote{Producing a high number of identical single photons (same direction, polarization, wavelength, ...) is a challenging task.} and modes employed limits the degradation of the entanglement and coherence due to interactions with the environment, and thus facilitates the physical implementation.

We demonstrate here the feasibility of these photonic-based QML models coupled to the classical DNN feature extractor for capturing the structure-property correlation patterns and chemistry in polymers. For this purpose, we have selected a classification task referring to a ground-state property such as the polymer gap. For ground-state properties, it suffices to model the polymer by using its monomer.

The outline of the paper is as follows. Section \ref{sec:II} introduces the polymer data, its featurization as well as the hybrid classical-quantum approach. We also give details on the classical DNN feature extractor, quantum models and linear QPCs. After providing the computational details in Section \ref{sec:Computational_information}, we discuss the results of the polymer classification in Section~\ref{sec:Results}. Finally, we conclude in Section \ref{sec:Conclusions}.  

\section{Polymer gaps classification: Hybrid classical-quantum approach} \label{sec:II}
\subsection{Polymer Data and Encoding}\label{sec:data_set}
As noted in section~\ref{Introduction}, high-quality databases for polymers %, yielded by polymer science and engineering, 
are rare and very often incomplete \cite{data_driven, ref31, ref32} \footnote{Non-formatted experimental polymer data is spread throughout journals, while public databases are rare.}. This sparsity can complicate the application of DL methods to property predictions, as the former require large amounts of data in order to fit the network parameters. In this study, we rely on computational studies for our data.  
\begin{figure*}[!htp]
        \centering
        \begin{subfigure}[b]{0.485\textwidth}
            \centering
            \includegraphics[width=\textwidth]{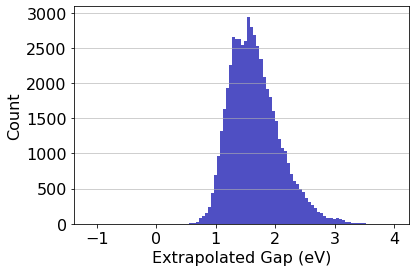}
            \caption[]%
            {{\small Distribution of the polymer extrapolated gaps (in eV).}}    
            \label{fig:data_set_distribution}
        \end{subfigure}
        \hfill
        \begin{subfigure}[b]{0.48\textwidth}  
            \centering 
            \includegraphics[width=\textwidth]{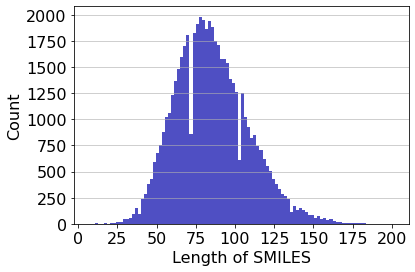}
            \caption[]%
            {{\small Distribution of the length of the monomer SMILES strings.}}
            \label{fig:lenght_smiles_distribution}
        \end{subfigure}
        \caption{Polymer dataset statistics}
    \end{figure*}
We use a dataset obtained from quantum chemistry calculations \cite{ref4, ref37} on monomers or oligomers based on Density Functional Theory (DFT) employing either B3LYP \cite{ref38, ref39} or CAM-B3LYP functionals \cite{ref40}.~The dataset is extracted from a database containing calculated optoelectronic properties of oligomers for organic photovoltaic (OPV) \cite{ref37} applications. Thus, it also contains electronic properties extrapolated to the polymer limit.~These include the extrapolated highest occupied molecular orbital $\varepsilon_{HOMO}$, lowest unoccupied molecular orbital $\varepsilon_{LUMO}$, gap $E_{g}$,\footnote{For a monomer, the gap is equal to its first excitation energy that, at first approximation, equals $\varepsilon_{LUMO}$ - $\varepsilon_{HOMO}$.} and the sum of the polymer $\varepsilon_{HOMO}$ and $E_{g}$. We focus here only on the polymer gap.     

In this work, each polymer species is represented by the SMILES of its monomer (or oligomer), sufficient when studying ground-state-related properties. To obtain a machine-readable form of the SMILES, we follow the work of Chen \textit{et al.} \cite{ref5}. They proposed a chemical language processing model based on monomer SMILES for predicting the glass transition temperature of polymers.~The model borrows encoding strategies from the field of Natural Language Processing (NLP), with details in appendix \ref{sec:Encode_smiles}. The encoding yields an integer-value vector for each SMILES in the dataset.~The encoded SMILES are one of the simplest feature representations of the chemical language, but this representation is not enough for extracting pertinent information about the underlying chemical structures. Moreover, it has been shown \cite{ref44} that this type of encoding alone leads to DNN models of poor accuracy.~The latter can largely be improved by combining the representation with a subsequent character embedding, as also proposed in Ref. \cite{ref5}.
The classical DNN embedding layer (see Fig. \ref{fig:lstm}) transforms each integer-value vector into a more compact and denser vector containing information about the chemical structure.
These vectors are fed into the next layers of a DNN which learn intricate correlations between the chemical structures of the polymers and their gap-derived classes. 

\subsection{Methods}\label{sec:methods}
\subsubsection{Classical data description}\label{sec:classical_data}
The initial dataset, yielded by the B3LYP-DFT calculations \cite{ref4}, contains 54335 unique polymers, each represented by the canonical SMILES\footnote{SMILES (Simplified Molecular Input Line Entry System) is a chemical notation that permits to represent a chemical structure in a computer-adapted numerical form.} string of its monomer and the extrapolated value of its gap.~The distribution of those values in the dataset is shown in Fig. \ref{fig:data_set_distribution}, with the majority of the polymer gaps centered between 1.0 and 2.6 eV.
Following the authors of Ref.~\cite{ref41}, we have further classified the polymer gaps into three classes according to spectroscopic energy ranges: VIS, NIR and mid-infrared (MIR) range. Such a classification is justified as the dataset contains multi-ring heterocycle compounds, conjugated polymers used in OPV-applications.~The NIR values range from 0.4 (excluded) to 1.6 eV, the VIS ones from 1.6 eV (excluded) to 4.0 eV, and finally the MIR extend from 0.025 eV to 0.4 eV. 
\begin{figure}[htb!]
    \centering
    \includegraphics[width=0.75\columnwidth]{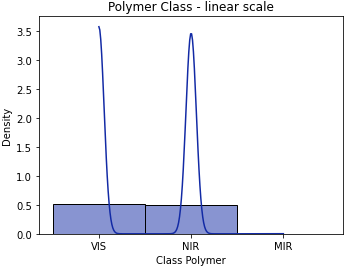}
    \caption{Density distributions and normalized counts of the polymer classes in the dataset of 54335 polymers.}
    \label{fig:classes_distribution}
\end{figure} 
As seen from the distribution of the polymer classes in Fig. \ref{fig:classes_distribution}, the dataset is largely dominated by classes VIS and NIR, and it is balanced with respect to them. In the following, we filter out the few polymers of class MIR, reducing the polymer studies to a binary classification problem. 

\begin{figure}[htb!]
    \centering
    \includegraphics[width=0.7\columnwidth]{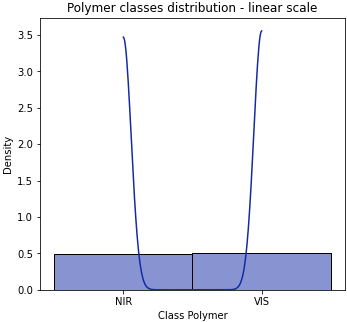}
    \caption{Density distributions and normalized counts of the polymer classes in the dataset of 52815 polymers.}
\label{fig:classes_distribution_2}
\end{figure}

The distribution of the SMILES strings length in the dataset is shown in Fig. \ref{fig:lenght_smiles_distribution}.~This distribution is not uniform, it is symmetric with a mean, a median and a maximum string length of 85, 83, and 201 characters, respectively\footnote{The asymmetry or skewness coefficient equals 0.4594. As a measure of asymmetry in a distribution, a positive skew means a longer tail to the right. The skew of a perfectly symmetric distribution equals 0.}.~Furthermore, 98 \% of the SMILES strings in the dataset have a length shorter than 140 characters.~To prepare the features for the development of the classical deep learning model in section \ref{sec:RNN}, we adopted a constant length for all SMILES sequences.~Based on the above observations, we chose a critical value $L_c$ of 140 characters: all SMILES sequences in the dataset shorter than $L_c$ were padded with zeros, while those longer than $L_c$ were removed.~After identifying\footnote{We used the standard deviation method on the normalized gaps distribution as well as boxplot visualisations.} and removing samples non-representative\footnote{samples
whose gap value differs significantly from those of the majority of samples} for our dataset,
we obtain a dataset of 52815 polymers and their extrapolated gaps.    
As seen in Fig. \ref{fig:classes_distribution_2}, the polymer classes distribution remains intact after the dataset transformations.

\subsubsection{Hybrid classical-quantum approach}\label{HCQ}
The hybrid classical-quantum approach consists of two building blocks: a feature extractor based on a classical DNN and a Noisy Intermediate-Scale Quantum (NISQ)-compatible binary classifier that utilizes either GKM, QE-RKS, or VQC implemented by linear QPCs \cite{ref42}.~This implementation is realized on Perceval \footnote{https://perceval.quandela.net/}, an open-source framework for programming photonic quantum computers \cite{ref36}.

The classical DNN, presented in section \ref{sec:RNN}, learns the chemical structure of the polymers in the dataset as well as the correlations between the chemistry and the polymers' gap-related classes. Whereas a purely classical binary classification task can be carried out, here we have used the DNN as a feature extractor. We have thus adapted its architecture to this purpose (see, section \ref{sec:RNN}). The extracted features containing pertinent chemical information have the form of 2D vectors. They are fed to the quantum classifier as input vectors. The classical DNN extractor can be viewed as an efficient data-encoding method that also reduces the dimension of the input data to the QPC implementing supervised binary classification model \cite{ref42}. The architecture of our hybrid approach bears some resemblance to that of the hybrid classical-quantum neural networks proposed in Ref. \cite{TransferLearning}, as it also transfers information from the classical to the quantum part of the network.

A simple example of the dimension reduction obtained in our polymer study is shown in Appendix \ref{sec:CS_DNN}.          
\subsubsection{Classical DNN feature extractor}\label{sec:RNN}
As discussed in section \ref{sec:data_set}, the polymer's repeat unit (monomer) is represented by a SMILES string that is viewed as a sequential data of characters. The SMILES is the input data to the classical DNN (Sec. \ref{HCQ}) after being pre-processed and encoded into a machine-readable form.~The architecture of the network is then chosen to process the chemical sequential data.~Here, we construct a network that contains Recurrent Neural Network (RNN) layers of bidirectional Long-Short-Term Memory (LSTM) type (see, Ref. \cite{ref43} for a review on RNN and LSTM.).~It is shown in Fig. \ref{fig:lstm} and discussed further in Appendix \ref{sec:CS_DNN}.

To train, validate, and test this chemical language processing model, we have randomly sampled 90 \% of the dataset, i.e., 47534 polymer structures. We note that the sampled dataset remains balanced with respect to the two polymer classes VIS and NIR. Details on the optimal hyperparameters of the network, activation functions, training settings including the dataset splitting, are provided in appendix \ref{sec:CS_DNN}.  

The final trained model is characterised with an accuracy on the testing set equal to 0.879 for the binary classification of the polymers. The classical DNN feature extractor is next obtained by removing the last three dense layers of the network after the training process. It is next applied to the remaining 10 \% of the initial dataset, i.e., 5281 polymer structures, to obtain input 2D vectors for the linear QPC. Note that this data sample is also balanced with respect to the two classes and thus representative of the initial dataset. The 2D vectors are the two-dimensional classical data in our binary classification problem. We chose two-dimensional data vectors to be able to run the simulations, implemented with Perceval, on a classical computer. 

\subsubsection{Gaussian kernel samplers and QE-RKS}\label{sec:GKM_RNN}
\begin{figure}
    \centering
\includegraphics[scale=0.6]{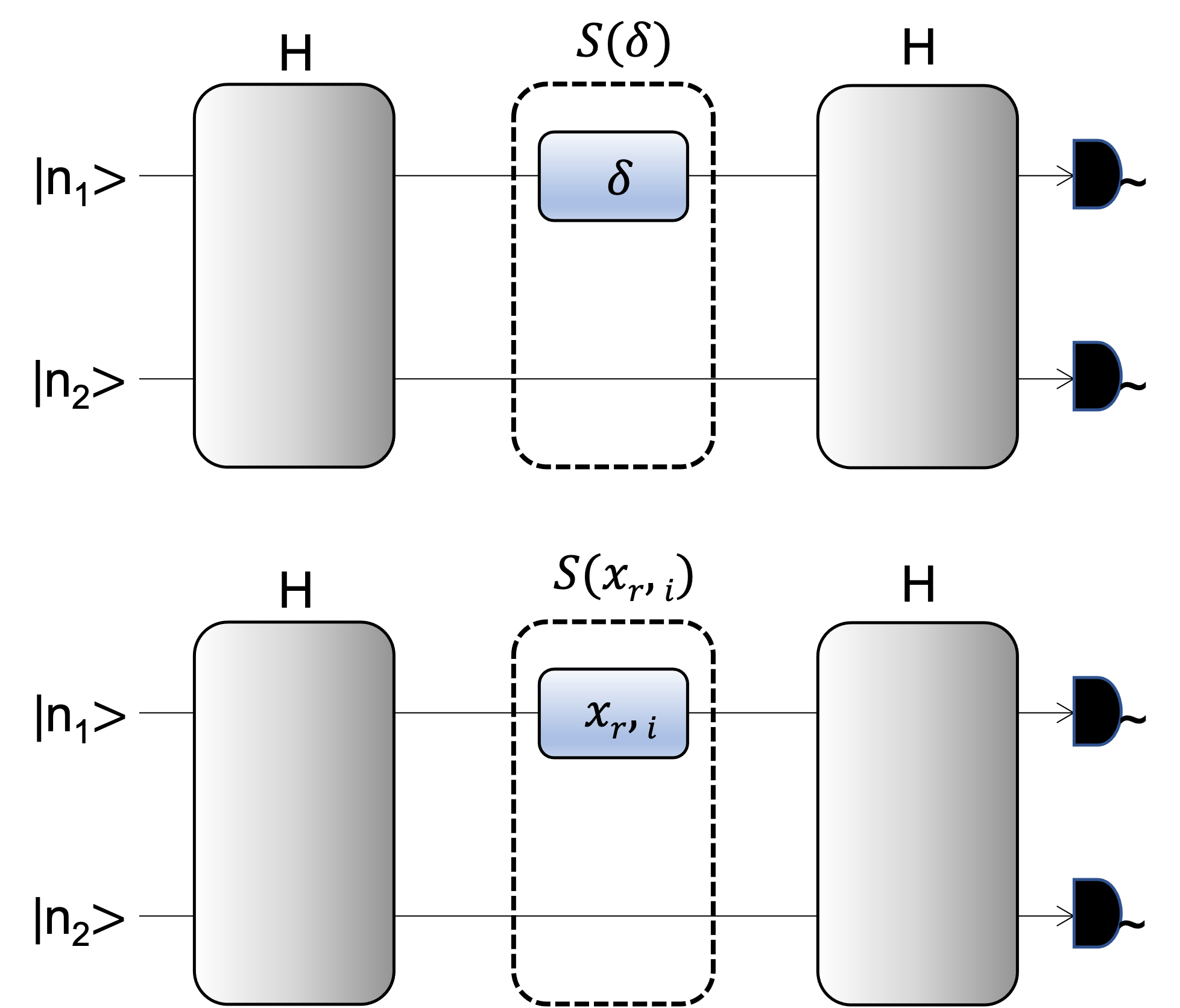}
    \caption{(\textit{illustration of Figs. 2b and 2c (arXiv) in Ref. \cite{ref42}}) Two-spatial modes linear QPCs for Gaussian kernel-based ML with photon number-resolving detectors. The top figure shows the QPC for the GKM, where a phase shifter encodes the distance between any two data points, $\mathbf{x}$ and $\mathbf{x}'$, $\delta = (\mathbf{x}- \mathbf{x}')^2$. The bottom figure illustrates the QPC for the QE-RKS approach that approximates the Gaussian kernel by sampling a set of randomized input features $x_{r, i} = \gamma(\mathbf{w}_r)\cdot \mathbf{x}_i + b_r$. In both figures, H is the Hadamard transform implemented by a 50-50 beam splitter.}
    \label{fig:LQC1}
\end{figure}
We first combine the classical DNN feature extractor with the approach from Ref. \cite{ref42} that employs fixed linear QPCs as Gaussian kernel samplers, avoiding thus the circuit optimisation. We have adopted here the two-spatial-mode circuits, presented in Fig. \ref{fig:LQC1}, that serve to implement a binary classification model using Gaussian kernels and photon number-resolving detectors. Note that the need of number-resolving detection complexifies the physical implementation. 
The QPC represents a Mach-Zehnder interferometer by construction, and its role is to approximate the Gaussian Ksernel. 
 
In a kernel approach, the 2D vectors, each representing a polymer (see, section \ref{sec:CS_DNN}), are casted into the feature space. The pairwise similarity between the data points in that higher dimensional space is measured by the kernel function $k(\mathbf{x}, \mathbf{x'})$. The latter is employed to calculate the decision boundary in the polymer classification task according to \cite{ref42, ref45}: 
\begin{align} \label{decision_boundary}
    f(\mathbf{x}) = \sum_{i=1}^{N}\beta_ik(\mathbf{x}_i, \mathbf{x}),
\end{align}
where $N$ is the number of the training data points, and $\mathbf{x}_i$ is the i$^{th}$ training 2D vector.
The similarities of a given sample $\bm{x}$ with the samples $\bm{x}_i$ are used to classify the sample $\bm{x}$.
From the equations $f(\bm{x}_j)$ = $y_j$ ($j \in \{ 1, N\}$), where $\mathbf{y}$ is the $N \times$1 vector of the training data labels, we obtain the linear system of equations 
\begin{align}
    K \bm{\beta} = \mathbf{y}, 
\end{align}
where K is the $N\times N$ kernel matrix K$_{ij}$=$k(\mathbf{x}_i,\mathbf{x}_j)$. After regularizing the system, we obtain
\begin{align}\label{LSEs}
    (K + \alpha\mathbf{I})\bm{\beta} = \mathbf{y}, 
\end{align}
where $\mathbf{I}$ and $\alpha$ are the $N$-dimensional identity matrix and the regularization parameter, respectively. The latter prevents the Gaussian kernel-based model from overfitting. We have thus reduced the optimization problem of finding the parameters $\beta$ to solving the linear system of equations in Eq. (\ref{LSEs}).
We adopt here the Gaussian kernel 
\begin{equation}
    k(\textbf{x}, \textbf{x}') = e^{-\frac{1}{2 \sigma^2} (\textbf{x} - \textbf{x}')^2}, 
\end{equation} 
but only consider the low-resolution case with $\sigma$ = 1.0. 

To obtain the approximate quantum kernel from the output of the two-mode QPC given by \cite{ref42}
\begin{eqnarray*}
f^{(n)}(\delta, \bm{\lambda}) = \sum_{ij} \lambda_{ij} |\langle n_i, n_j| \mathcal{U}(\delta)|n, 0\rangle|^2, \quad \text{where} \\ \quad \delta = (\mathbf{x}-\mathbf{x}')^2 \quad \text{and} \quad
\mathcal{U}(\delta) = \mathcal{H}\mathcal{S}(\delta)\mathcal{H},
\end{eqnarray*}
we have optimized the coefficients $\lambda_{ij}$ of the trainable quantum observable
\begin{eqnarray*}
\sum_{ij} \lambda_{ij} |n_i, n_j\rangle \langle n_i, n_j|,  \quad \text{$n_i$ + $n_j$ = $n$}, 
\end{eqnarray*}
with $n \coloneqq$ number of photons, by minimizing the quadratic loss function
\begin{equation}\label{loss_GM}
    \mathcal{L}(\delta, \bm{\lambda}) = \frac{1}{2 N} \sum_{i = 1}^N \left(f^{(n)}(\delta_i, \bm{\lambda}^{(\sigma)}) - e^{- \frac{\delta_i}{2 \sigma^2}} \right)^2, 
\end{equation}
where $\{\delta_i\}_{i = 1}^N$ is the training data. As depicted in Fig. \ref{fig:LQC1}, the unitary matrix $\mathcal{U}(\delta)$ contains two Hadamard transformations $\mathcal{H}$ implemented by 50–50 beamsplitters and a data embedding matrix $\mathcal{S}(\delta)$ encoded by a phase shifter. Note that no regularization term is needed in Eq. (\ref{loss_GM}), because the function to fit is regular and the training data points, drawn from the uniform distribution $U$(0, 3) are numerous and evenly spread over the interval (0,3). We exclude, therefore, any overfitting. 

We have also combined the classical DNN feature extractor with the Quantum-Enhanced RKS (QE-RKS) \cite{ref42}.  
The RKS algorithm approximates the researched kernel by using randomly-sampled feature maps. As the $n$-photon-derived quantum models are expressed in terms of Fourier series, the problem is reduced to a
sampling from random Fourier features. 
The RKS refers to a low-rank approximation of shift-invariant kernels. It consists of sampling a random subset of the kernels Fourier components.
The QE-RKS utilizes further the linear QPCs as a random feature sampler.~The random Fourier features that are $R$-dimensional vectors $\mathbf{z(x)}$ of randomized cosine functions
\begin{equation}\label{RandomFF}
\mathbf{z(x)} = \frac{1}{\sqrt{R}}
\begin{pmatrix}
    \sqrt{2} \cos (\gamma[\mathbf{w}_1 \cdot \mathbf{x} + b_1]) \\
    \vdots \\
    \sqrt{2} \cos (\gamma[\mathbf{w}_r \cdot \mathbf{x} + b_r]) \\
    \vdots \\
        \sqrt{2} \cos (\gamma[\mathbf{w}_R \cdot \mathbf{x} + b_R])
\end{pmatrix}
\end{equation}
with $\{\mathbf{x}\}$ being our 2D polymer data, approximate the Gaussian kernel as  
\begin{equation}
    e^{-\frac{\gamma^2}{2}(\mathbf{x}- \mathbf{x})^2}= k(\mathbf{x}, \mathbf{x'}) \approx \mathbf{z}(\mathbf{x})\cdot \mathbf{z}(\mathbf{x'}). 
\end{equation}
The 2D random vectors $\mathbf{w}_r$ are here drawn from a spherical Gaussian distribution, but other probability distributions such as the Laplacian and Cauchy ones can also be employed. Each can be used to approximate the other.
The random $b_r$ scalars are drawn uniformly from $[0, 2\pi]$. $\gamma$ controls the kernel resolution.~The solution to the decision boundary is then approximated as
\begin{align} \label{decision_boundary_rks}
    f(\mathbf{x}) \approx \underbrace{\sum_{i=1}^{N}\beta_i\mathbf{z}(\mathbf{x}_i)}_{\mathbf{c}} \cdot \mathbf{z}(\mathbf{x}),
\end{align}
where $\{(\mathbf{x}_i, y_i)\}_{i = 1}^{N}$ are the training data. The problem is then reduced to obtaining the optimal solution $\mathbf{c}_{opt}$ from a supervised linear model with respect to the $R$-dimensional (with ideally $R<<N$) random Fourier features $\mathbf{z}(\mathbf{x})$. As noted by Gan \textit{et al.}, the principal advantage of QE-RKS is that random Fourier features of different frequencies are sampled simultaneously by the linear QPC and used for constructing the kernel functions. In practice, the circuit is the same as that in the GKM. The difference comes from the randomized input encoding that is applied to each $\mathbf{x}_i$ and yields a set of randomized input features $x_{r, i}$. This circuit output is \cite{ref42}    
\begin{equation}
f^{(n)}(x_{r, i}, \lambda^{(k)}) = \sqrt{2} \cos (k\gamma[\mathbf{w}_r \cdot \mathbf{x}_i + b_r]),    
\end{equation}
with $k$ denoting the frequency of the cosine function. It has the structure of the random Fourier features in Eq. (\ref{RandomFF}). Gaussian kernels
\begin{equation*}
    e^{-\frac{k^2\gamma^2}{2}(\mathbf{x}- \mathbf{x})^2}= k(\mathbf{x}, \mathbf{x'}) 
\end{equation*}
with different resolutions $\frac{1}{k\gamma}$ can then be approximated by the QPC using different observables. In the following, the randomized input features $x_{r, i}$ are parameters for the phase shifters in the circuit.      
Note that the QE-RKS method only requires one photon. A threshold detection instead of a number-resolving one is, therefore, sufficient.  
\subsubsection{Variational Quantum Classifier}\label{sec:VQA}
\begin{figure}[!htbp]
\centering
\includegraphics[scale=0.6]{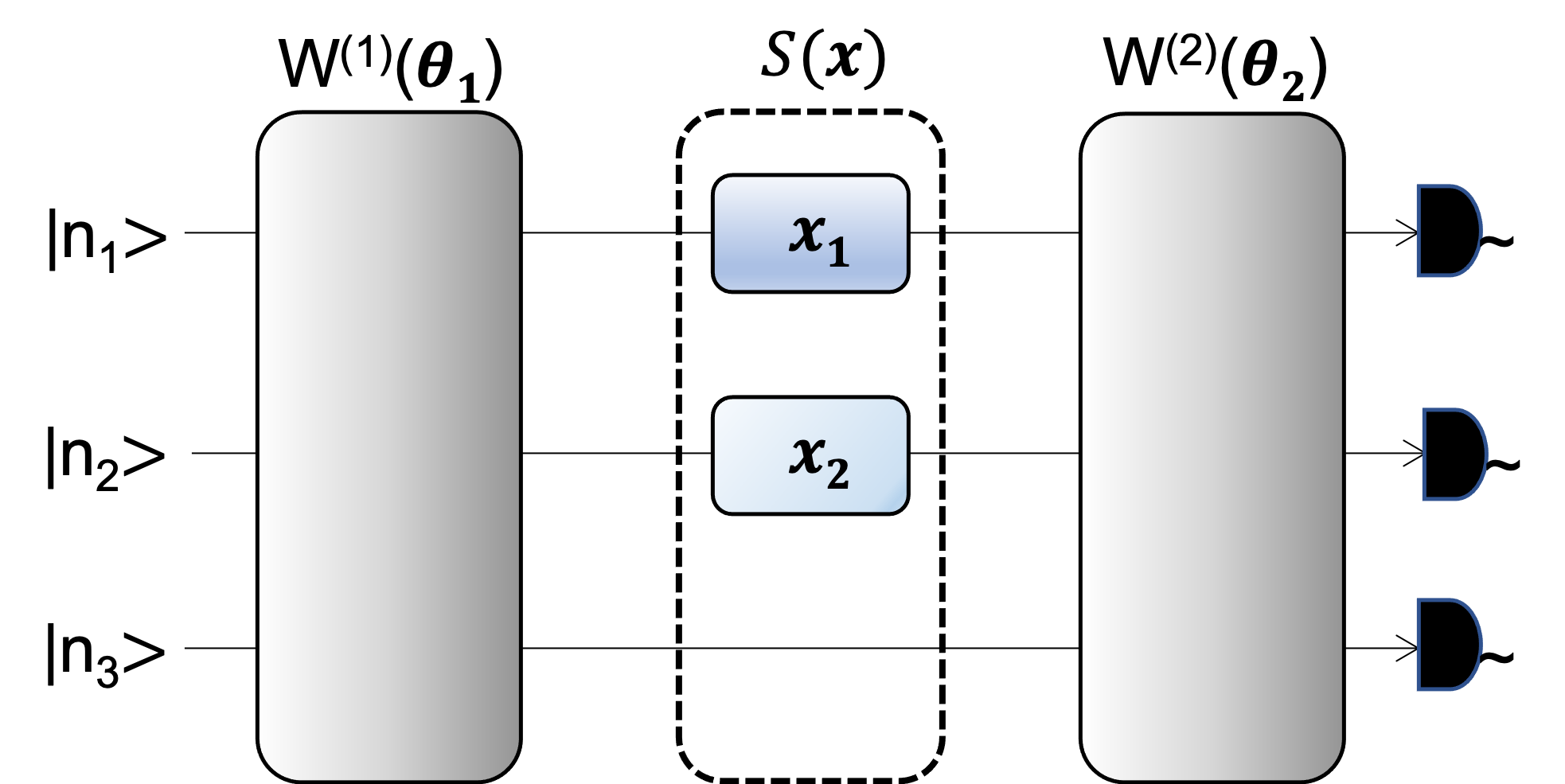}
\caption{(\textit{reproduction of Fig. 2a in Ref. \cite{ref42}}) Parameterized (by 
 $(\bm{\theta}_1, \bm{\theta}_2)$) three-spatial modes linear QPC for the variational quantum binary classifier. The circuit fits the Fourier series in the model. Each classical data feature is encoded by a tunable phase shifter [encoding layer $S(\bm{x})$]. W$^{(1)}(\bm{\theta}_1)$ and  W$^{(2)}(\bm{\theta}_2)$ are trainable beam splitter meshes.}
\label{fig:VQA}
\end{figure}
Finally, we have also considered the variational quantum classifier proposed in Ref. \cite{ref42}. The associated parameterized linear QPC with three spatial modes is depicted in Fig. \ref{fig:VQA}. 
The $n$-photon quantum model given by the QPC output reads \cite{ref42} 
\begin{eqnarray}
f^{(n)}(\bm{x}_l, \bm{\Theta}, \bm{\lambda}) &=& \sum_{ijk} \lambda_{ijk} \langle n_i, n_j, n_k| \mathcal{U}(\bm{\Theta}, \bm{x}_l)|n_1, n_2, n_3\rangle|^2, \nonumber \\ 
\mathcal{U}(\bm{\Theta}, \bm{x}_l) &=& \mathcal{W}^{(2)}(\bm{\theta}_2)\mathcal{S}(\bm{x}_l)\mathcal{W}^{(1)}(\bm{\theta}_1),
\end{eqnarray} 
$|n_1, n_2, n_3\rangle$ is the input $n$-photon Fock state with $n = n_1 + n_2 + n_3$ ($n_i$ the number of photons in mode $i$), and $\bm{\Theta} = (\bm{\theta}_1, \bm{\theta}_2)$ are the trainable circuit block’s parameters. $\mathcal{U}(\bm{\Theta}, \bm{x}_l)$ is a unitary transformation parameterized by $\bm{\Theta}$.~The QPC prepares a quantum state that is used to obtain the expectation value of the quantum observable parameterized by    $\bm{\lambda}$.  
The model is trained by employing the regularized squared loss function \cite{ref42}
\begin{equation}\label{loss_VQA}
    \mathcal{L}(\bm{\Theta}, \bm{\lambda}) = \frac{1}{2 N} \ \sum_{l = 1}^{N} (y_l - f^{(n)}(\bm{x}_l, \bm{\Theta}, \bm{\lambda}))^2 + \alpha \bm{\lambda} \cdot \bm{\lambda}, 
\end{equation}
where $\{y_l\}_{l=1}^{N}$ are the true data point labels, $N$ is the size of the training dataset $\{\bm{x}_l, y_l\}_{l=1}^{N}$ and $f^{(n)}(\bm{x}_l, \bm{\Theta}, \bm{\lambda})$ is the $n$-photon quantum model. $\alpha$ is the regularization weight.

The binary classification of the 2D polymer data is achieved by training the model to minimize the loss function in Eq. (\ref{loss_VQA}). We use a dual annealing minimization algorithm \cite{ref46}. As discussed in section \ref{sec:Computational_information}, the algorithm is selected as it permits to handle the strong non-convexity of the loss function. The decision boundary of the trained model is given by the circuit output
\begin{equation*}
f^{(n)}_{sign} (\bm{x})  =  sgn[f^{(n)}(\bm{x}, \bm{\Theta}_{opt}, \bm{\lambda}_{opt})],
\end{equation*}
with the optimized parameters for the circuit and observable $\bm{\Theta}_{opt}$ and $\bm{\lambda}_{opt}$, respectively.

In summary, the decision boundary is defined analogously in the three methods. In GKM and QE-RKS, the model, that is the circuit output, is determined by resolving equations of the form of Eq. (\ref{LSEs}). In VQC, the model is implicitly determined by minimizing the loss function. The class of the polymer data points is then determined by the model's sign.  

\section{Computational details}\label{sec:Computational_information}
To carry out the polymer binary classification, we have divided the 2D vectors dataset of 5281 structures into training, validation, and testing subsets in the ratio (80:10:10). The dataset was shuffled and stratified in order to ensure the representability of the two polymer classes VIS and NIR in the derived subsets. The classes VIS and NIR are further encoded as labels 1 and -1, respectively using a categorical encoding. Note that the 2D vector data is normalized by construction.   

For the studies based on GKM, the coefficients $\bm{\lambda}$ of the trainable quantum observable are optimized by minimizing the quadratic loss function in Eq. (\ref{loss_GM}). In contrast to Gan \textit{et al.} \cite{ref42} who employed BOBYQA \cite{BOBYQA}, we have used the basin-hopping algorithm\footnote{Basin-hopping is an algorithm that combines a global stepping algorithm with a local minimization at each step.} \cite{ref47} as implemented in SciPy \cite{ref_scipy}. Since our data is normalized, in order to train the quantum circuit from the top figure in Fig. \ref{fig:LQC1}, we draw the training data $\delta$ = $(\bm{x}- \bm{x}')^2$ from the uniform distribution $U$(0,3). This domain restriction is justified since the normalized 2D vectors live in the square $[-1,1]^2$. Furthermore, because of the high-dimensional Fock space on which the classical data is mapped, the components of the quantum observable are of $\sim 10^{6}$ orders of magnitude. The vector $\bm{\lambda}$ lives thus in a high-dimensional large-magnitude space.
As the loss function evolves very slowly with the change in $\bm{\lambda}$, the computed gradient tends to zero.   
In order to enable the minimization of the loss function, the magnitude of $\bm{\lambda}$ is brought down to the order of $10^{-3}$. The circuit input Fock state $|n, 0\rangle$, where $n$ is the number of photons in the first mode, is then multiplied by $10^{9}$ to recover the magnitude of the components of the quantum observable. In order to obtain $\bm{\lambda}_{opt}$ for $n$ = 1, 2, 4, 5, 6, 7, 8 or 10 (values we chose to study), the basin-hopping algorithm is employed for combinations of the two parameters: $niter$ = $\{1, ..., 20\}$ that is the number of basin-hopping iterations and $niter\_basin$ = $\{1, ..., 20\}$ that is the number of calls to the algorithm for a given $n$. For each $n$, the combination $(niter, niter\_basin)$ that yields for the loss function  either the lowest value or a value below $10^{-4}$ is retained. We note that at each iteration $niter\_basin$, a new random $\bm{\lambda}$ is sampled from an uniform distribution. We also note that the training of the circuit to approximate the Gaussian kernel $f^{(n)}(\delta, \bm{\lambda}^{(\sigma)}) \approx  e^{- \frac{\delta}{2 \sigma^2}}$ (here $\sigma = 1.0$) needs to be performed once, the $\mathcal{H}$ blocks do not need a reconfiguration if the training data is changed. After finding $\bm{\lambda}_{opt}$ for each $n$, we have performed a search for an optimal value of the regularisation coefficient $\alpha$ (see, Eq. (\ref{LSEs})). Since its value 
impacts the expressive power of the model (circuit output), we have tuned it for each $n$ separately. The $\alpha_{opt}$ values were found through trial and error. They are shown in Figs. \ref{fig:GaussianKernels} 
and \ref{fig:GaussianKernel_method_results_annex_1}.

With regard to the QE-RKS calculations, for the sake of simplicity, we have used $k$=1 and tuned the parameter $\gamma$ to the value of $0.1$. In contrast to the GKM, the linear QPC learns to approximate here the function $x \mapsto \sqrt{2} \cos{x}$. In analogy with Eq. (\ref{loss_GM}), we have set up a quadratic loss function and minimized it using the basin-hopping algorithm in order to obtain $\bm{\lambda}_{opt}$. The input Fock state $|n, 0\rangle$ for the circuit is here defined for $n$ = 1. We have also tested circuits with $n$ = 10. As expected, no advantage is observed with QE-RKS when increasing the value of $n$, using one photon suffices.      

In the VQC approach (section \ref{sec:VQA}), the classification of the data points is determined by the sign of the trained model (circuit output). Both observable and circuit parameters $\bm{\lambda}$ and $\bm{\Theta}$ are optimized while training the three-mode linear QPC in Fig. \ref{fig:VQA}. We have 
used the dual annealing optimization, a stochastic global optimization algorithm \cite{ref46}, as implemented in SciPy \cite{ref_scipy}. Experimental runs have shown that the loss function is strongly non-convex and hence, the gradient descent approach is not applicable. After testing different minimization procedures, we have found that the dual annealing is apt for this task. Due to its stochastic nature, the solution for $\bm{\lambda}$ may vary for different runs. Therefore, we have repeated the calculation 8 times for either case of $n$ = 1, $n$ = 3 or $n$ = 5 ($n$ being the number of input photons) and retained the result with the lowest loss function value ($< 0.28$) and highest score. 
All calculations are performed with a regularisation weight $\alpha$ = 0.0001. 
In an attempt to improve further the classification results, we also added an optimization of scaling and translation hyperparameters on a data subset unseen by the model.

\section{Results}\label{sec:Results} 
We have first applied the Gaussian kernel-based binary classification method for classifying the polymer data presented in sections \ref{sec:RNN}. 
In order to study the expressivity of the two-mode linear QPC, we have performed classifications with different input Fock states $|n, 0\rangle$, with $n$ = 1, 2, 4, 5, 6, 7, 8, 10 being the number of input photons in the $m$ = 2 spatial modes.

\onecolumngrid
\begin{widetext}
\begin{figure}[!htpb]
    \centering
    \includegraphics[width=0.8\textwidth]{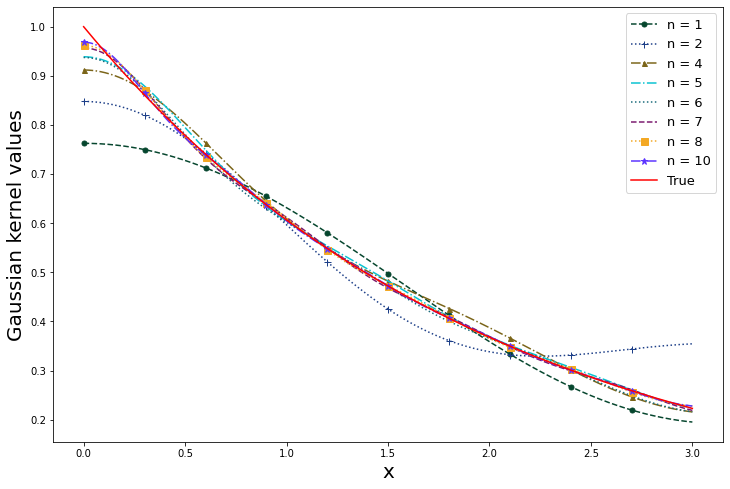}
    \caption{Approximating Gaussian kernel with a resolution $\sigma$ = 1.00 by employing the two-mode linear QPC from the top figure in Fig. \ref{fig:LQC1}. $\mathbf{x}$ denotes different values of $\delta = (\mathbf{x}- \mathbf{x}')^{2}$.
    }
    \label{fig:GaussianKernels}
\end{figure}
\end{widetext}

As discussed by Gan \textit{et al.} \cite{ref42}, low resolution kernels with $\sigma$ = 1.00 should be well-approximated by a circuit with a small number of input photons. We have plotted in Fig.~\ref{fig:GaussianKernels} the results from the training of the quantum kernel with circuits with different values of $n$. Indeed, increasing the number of input photons leads to more accurate approximate kernels. However, at least n = 4 photons are needed to fit the Gaussian kernel with $\sigma$ = 1.00 (red line in Fig.~\ref{fig:GaussianKernels}) as compared to n = 2 reported by Gan \textit{et al.}  
    \begin{figure*}[!htp]
        \centering
        \begin{subfigure}[b]{0.495\textwidth}
            \centering
            \includegraphics[width=\textwidth]{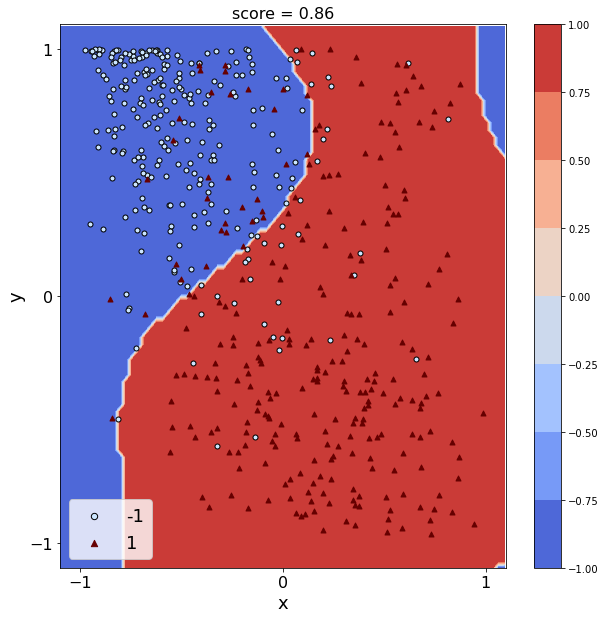}
            \caption[Number of photons: 1]%
            {{\small Number of photons: 1}}  
            \label{fig:num_1}
        \end{subfigure}
        \hfill
        \begin{subfigure}[b]{0.495\textwidth} 
            \centering 
            \includegraphics[width=\textwidth]{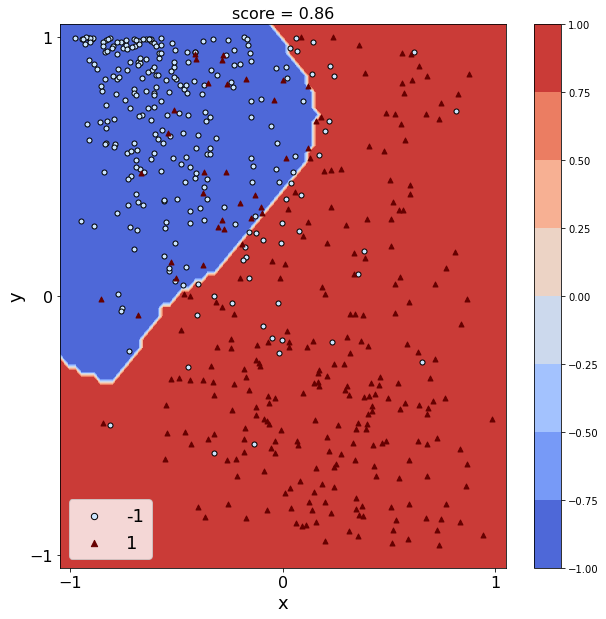}
            \caption[]%
            {{\small Number of photons: 4}}  
            \label{fig:num_4}
        \end{subfigure}
        \vskip\baselineskip
        \begin{subfigure}[b]{0.495\textwidth} 
            \centering 
            \includegraphics[width=\textwidth]{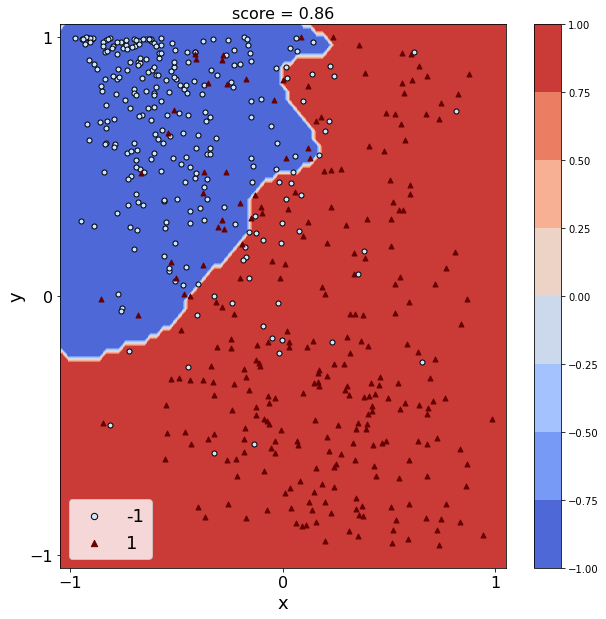}
            \caption[]%
            {{\small Number of photons: 5}}  
            \label{fig:num_5}
        \end{subfigure}
        \hfill
        \begin{subfigure}[b]{0.495\textwidth}  
            \centering 
            \includegraphics[width=\textwidth]{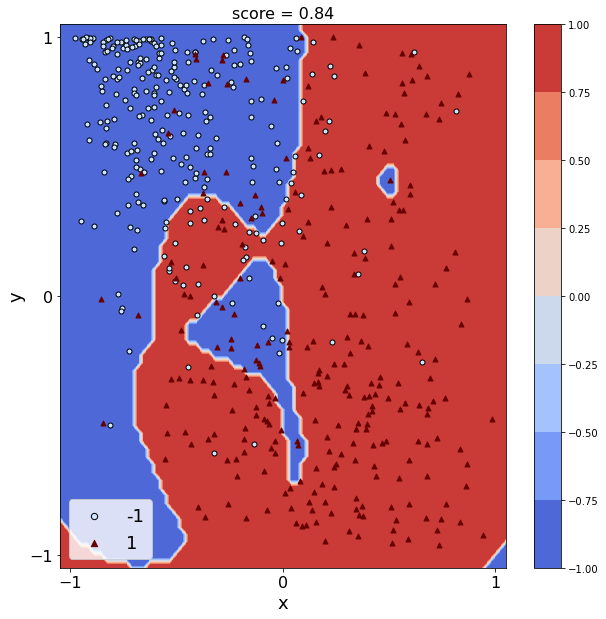}
            \caption[]%
            {{\small Number of photons: 10}} 
            \label{fig:num_10}
        \end{subfigure}
        \caption[ Gaussian Kernel Methods ]
        {\small Binary classification of polymers using the trained quantum kernels obtained with the two-mode linear QPC from the top figure in Fig. \ref{fig:LQC1} and with a different number of input photons $n$ = 1, 4, 5, 10. The regularisation parameter $\alpha$ equals 2.0, 3.0, 2.0, and 4.0, respectively. Test data points are shown in light-blue circles (class VIS [-1]) and dark-red triangles (class NIR [1]).
        }   \label{fig:GaussianKernel_method_results}
    \end{figure*}
In fact, the larger the fitting domain, the harder it is to fit the function and the higher Fourier components and hence, number of photons $n$ are needed. 

We have next employed the trained quantum kernels for classifying the polymer species. Figure \ref{fig:GaussianKernel_method_results} illustrates the classification (decision) boundaries obtained using the two-mode linear QPC with $n$ = 1, 4, 5, and 10 from Fig. \ref{fig:LQC1} (Figs. 2a and 2b in Ref. \cite{ref42}). Results for $n$ = 2, 6, 7, and 8 are shown for completeness in Fig. \ref{fig:GaussianKernel_method_results_annex_1}. 
Different values of the regularization parameter $\alpha$ were tested for each experiment with $n$ photons. The optimal values of $\alpha$ are reported in Fig.~\ref{fig:GaussianKernel_method_results} and reflect the complexity of the polymer dataset: larger values of $\alpha$ are likely needed in order to prevent the model from overfitting.
The score measuring the performance of each classifier is computed using the validation dataset.    
As expected from the work of Gan \textit{et al.} on artificially generated datasets, the classification boundaries become more complex with increasing the number of input photons which reflects the expressivity of the quantum circuit models. The trained quantum kernels approximate accurately the Gaussian one for n$\geq$4, but the performance of certain classifiers as measured by their validation scores declines for larger $n$ values. For n$\geq$7, the models score deteriorates
(see, the blue islands in Figs.~\ref{fig:GaussianKernel_method_results} and \ref{fig:GaussianKernel_method_results_annex_1}),
despite the larger regularisation factors employed. %The overfitting yields incorrectly classified data.         
   \begin{figure*}[!htp]
        \centering
        \begin{subfigure}[b]{0.495\textwidth}
            \centering
            \includegraphics[width=\textwidth]{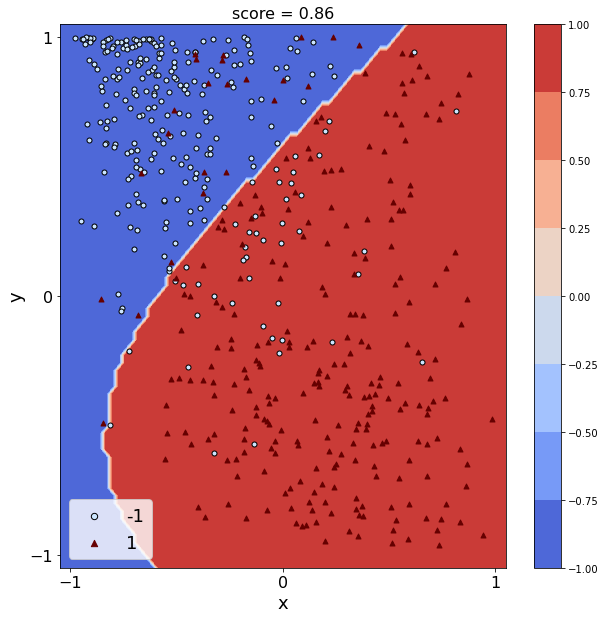}
            \caption[Random Fourier features by  
        sampling from a $\chi$-square distribution.]%
            {{\small  $\chi^2$ distribution, $R$ = 3}}  
            \label{fig:rks_chi_1photon}
        \end{subfigure}
        \hfill
        \begin{subfigure}[b]{0.495\textwidth} 
            \centering 
            \includegraphics[width=\textwidth]{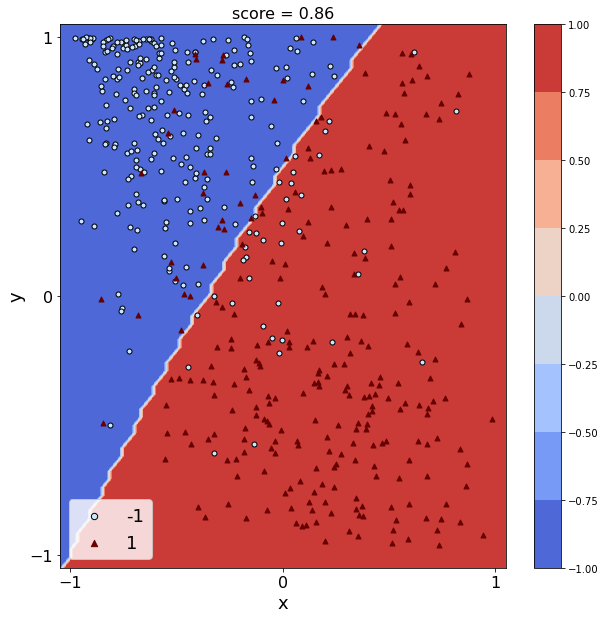}
            \caption[]%
            {{\small normal distribution, $R$ = 3}}  
            \label{fig:rks_normal_1photon}
        \end{subfigure}
                \hfill
        \begin{subfigure}[b]{0.495\textwidth} 
            \centering 
            \includegraphics[width=\textwidth]{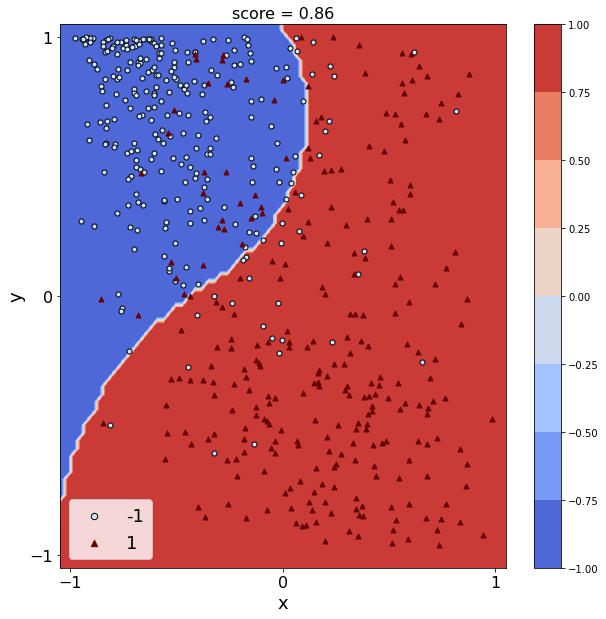}
            \caption[]%
            {{\small  $\chi^2$ distribution, $R$ = 100 }}  
            \label{fig:rks_chi_100_1photon}
        \end{subfigure}
                \hfill
        \begin{subfigure}[b]{0.495\textwidth} 
            \centering 
            \includegraphics[width=\textwidth]{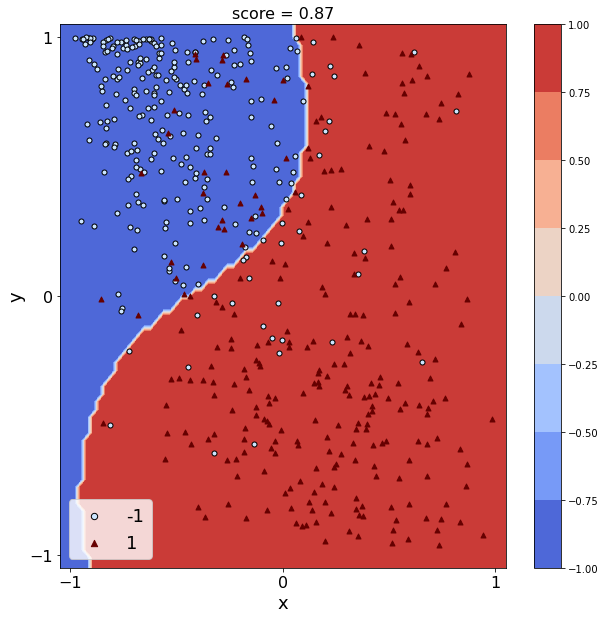}
            \caption[]%
            {{\small  normal distribution, $R$ = 100 }}  
            \label{fig:rks_normal_100_1photon}
        \end{subfigure}
        \caption[ QE-RKS Methods ]
        {\small Binary classification of polymers using the QE-RKS implemented by the two-mode linear QPC shown in the bottom figure of Fig. \ref{fig:LQC1}. The input Fock state is $|1, 0\rangle$. No regularization weight is applied.
        % The random seed was chosen to be 12345. 
        The random Fourier features are obtained by  
        sampling from either a normal or $\chi^2$ distribution.
        Test data points are shown in light-blue circles (class VIS [-1]) and dark-red triangles (class NIR [1]). 
        }
\label{fig:GaussianKernel_method_results_RKS}
    \end{figure*}   
The trained quantum kernel obtained with one input photon
poorly approximates the Gaussian kernel.
Yet, the associated classifier's score for the polymer dataset equals 0.86, signifying a satisfactory performance  
 (see, Fig. \ref{fig:num_1}).
The polymer dataset has overlapping space regions, where polymers of both VIS and NIR classes are present.~However, either region is populated predominantly by only one of the two classes. The associated point clouds distribution is such that the classification boundary can already be learned using circuits with a small number of input photons, $n$ = 1, 2. Note that when $n$ = 1, the kernel reduces to a cosine kernel, that is also a valid kernel. The contour plots in Fig. \ref{fig:GaussianKernel_method_results} show also some non-linearity of the decision boundary even for $n$ = 1, 2.

Coming back to the trained quantum kernels, we note that while larger $n$ values increase the expressivity of the linear QPC, which then yields more accurate kernels, they can also lead to a deterioration of the models scores, see Figs. \ref{fig:num_10} and \ref{fig:GaussianKernel_method_results_annex_1}.
The quantum kernel matrix that follows the true Gaussian kernel matrix (here for $\sigma$ = 1.0) approximates the inner product of the vectors in the high-dimensional feature space.~The approximation precision is, however, insufficient to obtain an accurate binary classifier. Therefore, the regularization term in Eq. (\ref{LSEs}) has an important role while optimizing the coefficients $\beta$ in Eq. (\ref{decision_boundary}). 
   \begin{figure*}[!htp]
        \centering
        \begin{subfigure}[b]{0.495\textwidth} 
            \centering 
            \includegraphics[width=\textwidth]{Figures/RKS_R3_normal_1photon.png}
            \caption[]%
            {{\small $R$ = 3}}  
            \label{fig:rks_r_3_1photon}
        \end{subfigure}
        \hfill
        \begin{subfigure}[b]{0.495\textwidth} 
            \centering 
            \includegraphics[width=\textwidth]{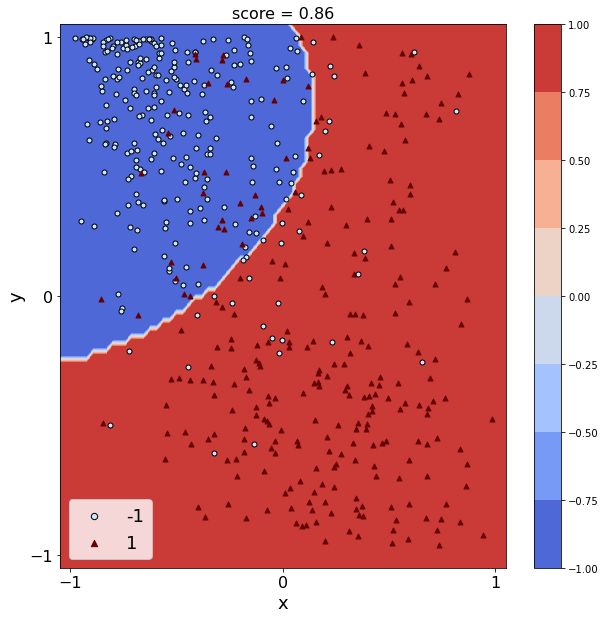}
            \caption[]%
            {{\small $R$ = 10 }}  
            \label{fig:rks_r_10_1photon}
        \end{subfigure}
        \hfill
        \begin{subfigure}[b]{0.495\textwidth} 
            \centering 
            \includegraphics[width=\textwidth]{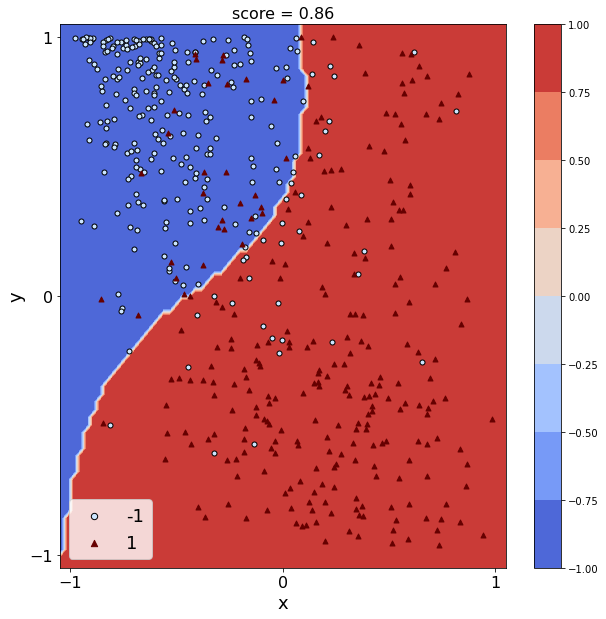}
            \caption[]%
            {{\small $R$ = 20 }}  
            \label{fig:rks_r_10_1photon}
        \end{subfigure}
        \hfill
        \begin{subfigure}[b]{0.495\textwidth} 
            \centering 
            \includegraphics[width=\textwidth]{Figures/RKS_R100_normal_1photon.png}
            \caption[]%
            {{\small $R$ = 100 }}  
            \label{fig:rks_r_100_1photon}
        \end{subfigure}
        \caption[ QE-RKS Methods 2 ]
        {\small Binary classification of polymers using the QE-RKS implemented by the two-mode linear QPC shown in the bottom figure of Fig. \ref{fig:LQC1}. The input Fock state is $|1, 0\rangle$. No regularization weight is applied. The random Fourier features are obtained by  
        sampling from a normal distribution. 
        Test data points are shown in light-blue circles (class VIS [-1]) and dark-red triangles (class NIR [1]). 
        }
\label{fig:GaussianKernel_method_results_RKS_2}
    \end{figure*} 
    
In the next experiments, we employ the QE-RKS model proposed by Gan \textit{et al.} \cite{ref42} to the classification of the 2D polymer data. The circuit  exploited in these calculations is shown in Fig. \ref{fig:LQC1}, bottom figure. This linear QPC is the random feature sampler.    
We have first compared the performance of the QE-RKS for the polymer data employing circuits with one input photon, i.e., input Fock states $|1, 0\rangle$ and random Fourier features of dimension $R$ = 3, but sampled from either normal or $\chi^2$ distributions. % We recall that the seed is fixed at value 12345. 
The results are shown in Fig. \ref{fig:GaussianKernel_method_results_RKS}.
As expected, for small values of $R$ 
% (even for a fixed seed)
the decision boundaries differ depending on the sampling distributions.~Yet the performances as measured by the score metrics are comparable.~As seen from Figs. \ref{fig:rks_chi_100_1photon} and \ref{fig:rks_normal_100_1photon}, the impact of the sampling distribution on the decision boundary shape decreases when increasing the value of $R$. When $R$ increases, the random Fourier features of higher dimension should approximate better the kernels and hence, improve the binary classifier. Note also that the performance of the QE-RKS method becomes more uniform when $R$ increases because of the local uniform convergence of the model (see, section \ref{QE-RKS_annex}).~As we approximate kernels with a low resolution ($k$ = 1 and $\gamma$ = 0.1), the decision boundaries are practically noiseless.~The performance of the models for $R \geq$ 20 are comparable (Fig. \ref{fig:GaussianKernel_method_results_RKS_2}). We note that the QE-RKS models (with either $n$=1 or $n$=10) obtained for the polymer data perform better than the GKM model employing a circuit with 10 input photons.~Whereas other datasets and input Fock states might show different trends, QE-RKS delivers classification models which performance is comparable to that of GKM models (see, also results on artificial datasets in Ref. \cite{ref42}).~This is important since finding the coefficients $\bm{c}_{opt}$ in the RKS algorithm scales as $O(R^3)$ while
the GKM scales as $O(N^2)$, where $N$ is the number of the training data points \cite{ref42}. This is the case because the classical data is mapped onto a lower-dimensional space of randomized $R$-dimensional Fourier features.   
With $R_{max}$ = 100 and $N \approx$ 5000 ($R<<N$), we have obtained a classification model at a lower computational cost and with a comparable accuracy. 

We have also carried out experiments with input Fock states $|10, 0\rangle$. Results are shown in Figs. \ref{fig:GaussianKernel_method_results_RKS_10photons} and \ref{fig:GaussianKernel_method_results_RKS_2_10photons}. As expected, the performance of the obtained models is comparable to that of the one-photon-derived models.    

% VQC
   \begin{figure*}[!htpb]
        \centering
        \begin{subfigure}[b]{0.495\textwidth}
            \centering
            \includegraphics[width=\textwidth]{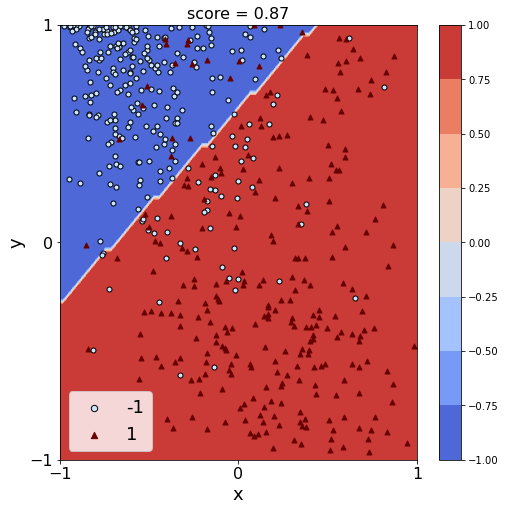}
            \caption[Input Fock state are $|100\rangle$]%
            {{\small Number of photons: 1; 
            Input Fock state $|100\rangle$ }}  
            \label{fig:vqa_n1}
        \end{subfigure}
        \hfill
        \begin{subfigure}[b]{0.495\textwidth} 
            \centering 
            \includegraphics[width=\textwidth]{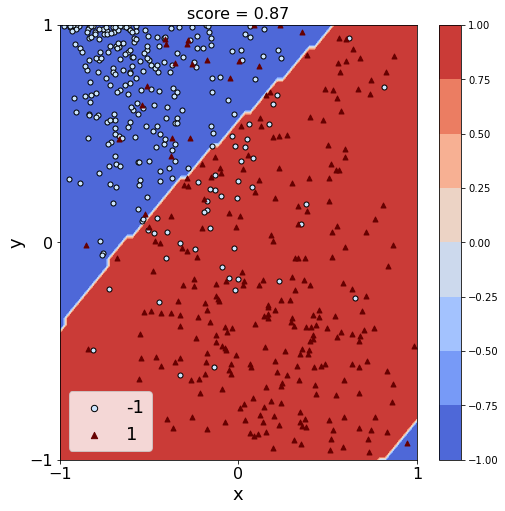}
            \caption[]%
            {{\small Number of photons: 3;}
            {\small Input Fock state $|111\rangle$}
            }  
            \label{fig:vqa_n2}
        \end{subfigure}
         \hfill
        \begin{subfigure}[b]{0.495\textwidth} 
            \centering 
            \includegraphics[width=\textwidth]{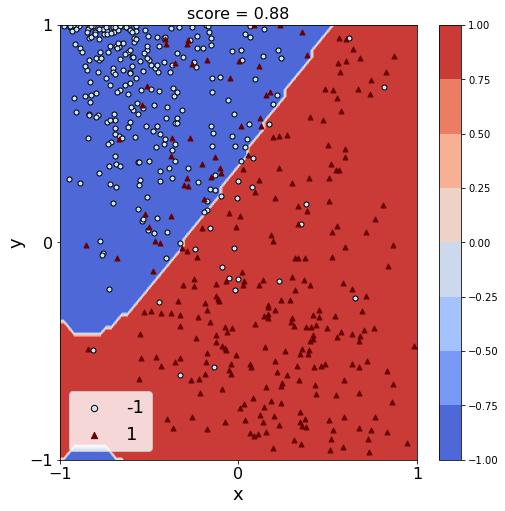}
            \caption[]%
            {{\small Number of photons: 5;}
            {\small Input Fock state $|221\rangle$}
            }  
            \label{fig:vqa_n3}
        \end{subfigure}
        \caption[ VQA Methods ]
        {\small Binary classification of polymers using the VQC implemented with the help of the three-mode linear QPC in Fig. \ref{fig:VQA}. Three input Fock states $|100\rangle$, $|111\rangle$, and $|221\rangle$ are used. The regularization weight $\alpha$ = 0.0001. Test data points are shown in light-blue circles (class VIS [-1]) and dark-red triangles (class NIR [1]). 
        }
\label{fig:GaussianKernel_method_results_vqa}
    \end{figure*}
Finally, we tested the performance of the VQC implemented by the trainable linear QPC depicted in Fig. \ref{fig:VQA}.~It consists of three spatial modes for fitting the $n$-photon quantum model as well as one encoding phase shifter per classical data vector component. In contrast to the two-mode linear QPC implementing the GKM or QE-RKS, here the circuit and quantum observable parameters are both trained explicitly. Fig. \ref{fig:GaussianKernel_method_results_vqa} summarizes the results for three different input Fock states $|100\rangle$, $|111\rangle$ and $|221\rangle$ corresponding to a number of input photons $n$ = 1, $n$ = 3 and $n$ = 5, respectively. As explained in section \ref{sec:Computational_information}, we have performed each experiment for either $n$ = 1, $n$ = 3 or $n$ = 5 eight times because of the stochastic nature of the quantum annealing  algorithm that may result into somewhat different values for the parameters $\bm{\lambda}$. Results having the lowest loss function value ($<0.28$) and highest score are retained for each $n$. As in the case of the GKM, increasing the number of input photons produces more complex decision boundaries (Fig. \ref{fig:vqa_n3}) since the expressive power of the quantum models increases as well. For our polymer dataset the classification model obtained with $n$ = 1 is already able to classify the samples satisfactory as measured by the model's score. Finally, we note that the optimization of the scaling and translation hyperparameters (see, Sec. \ref{sec:Computational_information}) brought no significant improvement in the results.                   
\section{Conclusion}\label{sec:Conclusions}
In this work, we leverage three quantum machine learning models \cite{ref42}, the GKM, the QE-RKS, and the VQC 
for the classification of polymer data. 
 The models exploit a photonic-based data-embedding scheme that consists of mapping the classical data points into the high-dimensional Fock space \cite{ref42}.~They are implemented using QPCs. Here, we combine them further with a classical DNN feature extractor that extracts information about the polymer structures, reducing, in addition, the classical data 139-dimensional vectors to 2-dimensional vectors.~The latter are the input data fed to the QPCs. Numerical representations for chemical species are often high-dimensional vectors whose encoding would require an important number of logical qubits and photons, respectively.~This hybrid classical-quantum strategy permits us to develop accurate quantum classifiers by leveraging Fock states with only \textit{a few} photons.~This is possible because of the high-dimensionality of the Fock basis states space.~Provided the current resource restrictions (NISQ architectures), this is important as it enables the transfer of the simulation to the quantum hardware. The focus of this study is not to obtain a quantum advantage, but rather demonstrate the feasibility of photonic quantum computing for classifying complex chemical structures such as polymers.

The hybrid classical-quantum approach has been applied to distinguish two polymer classes, VIS and NIR. The classes were determined based on the polymer gaps size.~The performance of the three hybrid classical-quantum models is found to be similar.~Their accuracy as measured by the model score metrics 0.86 -- 0.88 is comparable to that of our classical DNN model (score 0.87).

Whereas the three QML models were only applied to artificial datasets with some designed complexity in Ref. \cite{ref42}, our results demonstrate the potential of the hybrid approach to treat real data with a chemically-determined complexity.~More complex datasets should also be at its reach.

\section{Acknowledgements}
T. H. would like to acknowledge École Polytechnique for support. The authors would like to thank Alexia Salavrakos, Jason Mueller, Pierre-Emmanuel and Boris Bourdoncle for their constructive feedback.  

\bibliography{qcpolymers}

\appendix
\section{Encoding the monomers SMILES}\label{sec:Encode_smiles}
Following the encoding strategy from Ref. \cite{ref5}, each monomer SMILES is encoded at the character level. A dictionary of the unique characters in the SMILES corpus of the dataset is constructed, with each character being encoded by using its position in the dictionary. These positions range from 0 to the total number of unique characters minus 1 (0-based indexing). For the dataset here, this dictionary is given by
\begin{lstlisting}[language=python]
    {'#':0, '%':1, '(':2, ')':3, '-':4, 
    '.': 5, '/': 6, '0': 7, '1': 8, 
    '2': 9, '3': 10, '4': 11, '5': 12, 
    '6': 13, '7': 14, '8': 15, '9': 16, 
    '=': 17, '@': 18, 'C': 19, 'F': 20, 
    'H': 21, 'N': 22, 'O': 23, 'P': 24, 
    'S': 25, '[': 26, '\\': 27, ']': 28, 
    'c': 29, 'i': 30, 'n': 31, 'o': 32,
    's': 33}
\end{lstlisting}
and it contains 34 unique entries.~Each SMILES is represented by a vector which components are the index numbers of the SMILES characters in the above dictionary. For example, for the first three SMILES in the dataset, the corresponding vectors are 
\begin{align*}
v_1 = [19, 29,  8, ...,  0,  0,  0] \\
v_2 = [19, 26, 25, ...,  0,  0,  0] \\
v_3 = [19, 31,  8, ...,  0,  0,  0]
\end{align*}
signifying the following SMILES strings "Cc1...", "C[S...", "Cn1...", respectively.
All zeroes at the end have no special meaning and are only used
to pad the sequences to a fixed length of \textbf{139} characters.

\section{Computational settings for the classical DNN model}\label{sec:CS_DNN}

\begin{widetext}

\begin{figure}[!htbp]
    \centering
    \includegraphics[width=\textwidth]{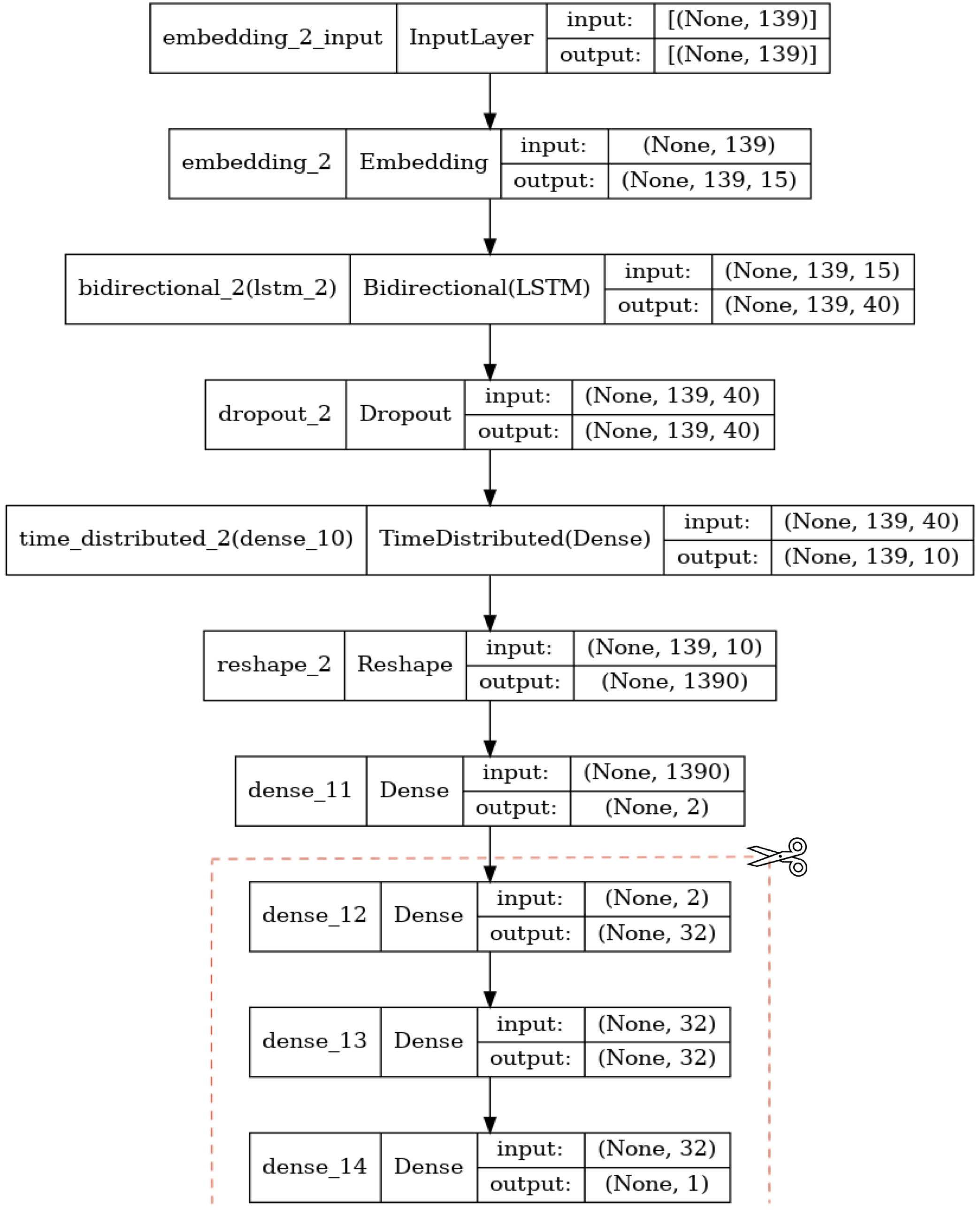}
    \caption{Classical DNN (from top to bottom: first layer [InputLayer] $\xrightarrow{}$ last layer [Dense]) Each layer is depicted by a rectangular containing information about the layer's type (Input, Embedding, LSTM, Dropout, Dense, ...) and the data dimensions at the input and output of the layer. For example, (None, 139, 40) allows a variable batch size (denoted by None), each sample in the batch having the shape (139, 40). The classical DNN feature extractor is obtained by discarding the last 3 Dense layers of the trained DNN model.}
    \label{fig:lstm}
\end{figure}

\end{widetext}

The classical DNN architecture is depicted in Fig.~\ref{fig:lstm}. 

An embedding layer is the first layer in the model (see, Sec. \ref{sec:data_set}).
The embedding layer is a dense representation in which similar chemical species have a similar encoding. The embedding layer is followed by a Bidirectional RNN-LSTM layer, a Time Distributed Dense layer, three Dense layers, and a final Dense layer for the output of the DNN. A dropout layer with a default dropout value of 0.0 is also added for a possible regularization of the network.
The two hyperparameters, the number of LSTM units in the Bidirectional LSTM layer, also used to define the number of units in the Time Distributed Dense layer, and the embedding layer output dimensions have the optimal values of 20 and 15, respectively. The activation functions are the ReLu functions for most Dense layers except for the last Dense layer, where a sigmoid is used. Also, the Dense layer positioned before the last three layers is equipped with a tanh activation function in order to ensure that the 2D vector values remain within the domain [-1., 1.].  

The classical DNN feature extractor is obtained by discarding the last 3 Dense layers of the trained DNN. It is then applied to the dataset of 5281 polymers to derive the 2D data vectors fed to the linear QPC.  Example of the dimension reduction obtained with the classical DNN feature extractor is shown below:
\[
\begin{bmatrix}
[19,  0, 19, ...,  0,  0,  0],\\
[19, 19,  2, ...,  0,  0,  0],\\
[19, 23, 19, ...,  0,  0,  0],\\
..., \\ 
[19, 23, 29, ...,  0,  0,  0],\\
[19, 22,  8, ...,  0,  0,  0],\\
[19, 31,  8, ...,  0,  0,  0]
\end{bmatrix} 
\longrightarrow 
\begin{bmatrix}
       [-0.88028246,  0.45909798],\\
       [ 0.7486655 ,  0.4901661 ], \\
       [ 0.9423334 , -0.8584387 ], \\
       ..., \\
       [ 0.05152836, -0.22515082], \\
       [ 0.8353953 , -0.8803364 ], \\
       [-0.49917743, -0.34748328] \\
\end{bmatrix}
\]
The 2D vectors (right) resulting from that reduction are, each, a compact representation of the polymer's chemical structure and its relation with a given class.  
    \begin{figure*}[!htpb]
        \centering
        \begin{subfigure}[b]{0.5\textwidth}
            \centering
            \includegraphics[width=\textwidth]{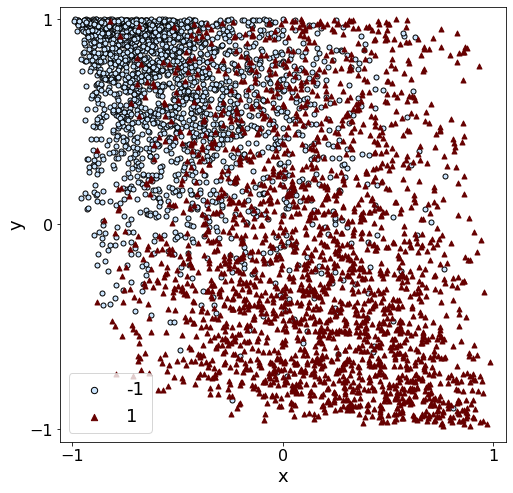}
            \caption[Training set]%
            {{\small Training dataset : 4276 2D vectors}}    
            \label{fig:}
        \end{subfigure}
        \hfill
        \begin{subfigure}[b]{0.49\textwidth}  
            \centering 
            \includegraphics[width=\textwidth]{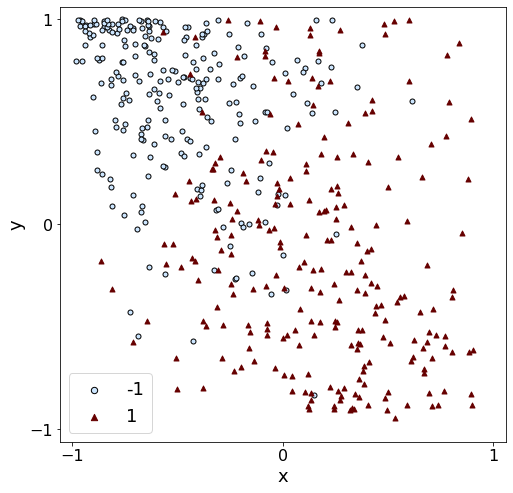}
            \caption[]%
            {{\small Validation dataset: 476 2D vectors}}    
            \label{fig:}
        \end{subfigure}
        \vskip\baselineskip
        \begin{subfigure}[b]{0.49\textwidth}   
            \centering 
            \includegraphics[width=\textwidth]{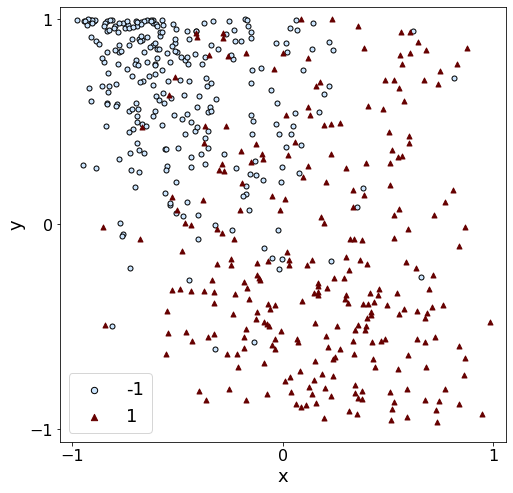}
            \caption[]%
            {{\small Testing dataset: 529 2D vectors}} 
            \label{fig:}
        \end{subfigure}
        \caption[Illustration of the polymer 2D dataset used as an input data to the linear QPC.]
        {\small Illustration of the polymer 2D dataset used as an input data to the linear QPC. The 2D vectors are yield by the classical DNN feature extractor.} 
        \label{fig:2D_input_data}
    \end{figure*}
The dataset of 2D vectors
is next divided into a training, validation, and testing subsets in the ratio (80:10:10). The resulting subsets are shown in Fig. \ref{fig:2D_input_data}.
\newpage

\section{GKM, QE-RKS and the corresponding circuits}

This section provides some elements on the GKM and QE-RKS. An intuitive description of the employed linear quantum circuit (LQC) is followed by some theoretical insights on the classical RKS method and its quantum enhancement.

\subsection{Understanding the circuit}

\begin{center}
    \begin{figure}[htp!]
      \centering
      \includegraphics[width=0.37\linewidth]{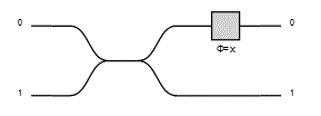}
\caption{Two-mode LQC with a phase shifter.}
      \label{fig:inter}
    \end{figure}
\end{center}

Any LQC can be expressed in terms of beam splitters and phase shifters \cite{generic}. We discuss, therefore, circuits containing only these optical elements. The phase shifter is the only one-mode component. Hence, we cannot use a one-mode circuit to express the cosine or Gaussian kernel. The circuit should contain at least two modes, i.e., $m$ = 2 (Fig. \ref{fig:inter}).
In order to prepare the input state $\ket{\text{n}, 0}$, where $n$ is the number of input photons, we apply a beam splitter followed by a phase shifter. If we apply a phase shifter $\phi$ to the first mode in Fig. \ref{fig:inter}, we obtain the following circuit unitary:
\begin{equation*}
    \begin{pmatrix}
    \frac{e^{i \phi}}{\sqrt{2}} & \frac{ie^{i \phi}}{\sqrt{2}}\\
    \frac{i}{\sqrt{2}} & \frac{1}{\sqrt{2}}
    \end{pmatrix}
\end{equation*}
After measuring the circuit we obtain constant probabilities.~Therefore, we add a beam splitter as shown in Fig. \ref{fig:LQC1} (bottom) to compute the cosine kitchen sink \cite{ref42}. The following unitary matrix is obtained:
\begin{equation*}
    \begin{pmatrix}
    \frac{e^{i \phi} - 1}{2} & \frac{i (e^{i \phi} + 1)}{2}\\
    \frac{i (e^{i \phi} + 1)}{2} & -\frac{e^{i \phi} - 1}{2}
    \end{pmatrix}
\end{equation*}
The measured probabilities can be written as 
\begin{equation} \label{prob}
    |\langle n-m, m|{\mathcal{U}(\phi)}|{n, 0}\rangle|^2 = \binom{n}{m} \cos^{2 m}\left(\frac{\phi}{2}\right) \sin^{2 (n - m)}\left(\frac{\phi}{2}\right),
\end{equation}
where $\mathcal{U}(\phi)$ is a unitary matrix. 
Hence, we can express any Fourier transform by linearizing Eq. (\ref{prob}). This justifies the definition of the circuit in Fig. \ref{fig:LQC1}. Moreover, we see that the expressivity of the circuit depends mainly on the number of photons employed.

\subsection{QE-RKS method}\label{QE-RKS_annex}
\subsubsection{Classical RKS}

Kernel methods rely on transforming a nonlinear classification problem into a linear problem in a high-dimensional space. As explained in Ref. \cite{maria}, these methods offer a much larger expressivity than the conventional variational methods. However, they also require the computation of kernel functions. In general, a variational approach is needed in order to define a LQC that computes the desired kernel. This is the strategy used in Ref. \cite{ref42} for computing the Gaussian kernel with a boson sampler.

Rahimi and Recht \cite{rahimi-recht1, rahimi-recht2} proposed to replace the costly optimization process by a randomization. In particular, the model becomes a linear sum of simple nonlinear functions that have random parameters: 
\begin{equation} \label{model}
    f(\textbf{x}) = \frac{1}{\text{R}} \sum_{i = 1}^R c_i z_i(\textbf{x)},
\end{equation} 
where 
\begin{equation} \label{cos-sink}
    z_i(\textbf{x}) = \cos(\omega_i \textbf{x}_i + \text{b}_i).
\end{equation}
These functions are called kitchen sinks, hence the name of the method. The weights of the kitchen sinks, $c_i$, in the linear sum are optimized. Such an optimization is a linear problem.

Usual kitchen sinks are the cosine, sine and sign functions. Using the cosine kitchen sink (see, Eqs. (\ref{model}) and (\ref{cos-sink})), Rahimi and Recht \cite{rahimi-recht0} have shown that the frequency parameter follows the probability distribution defined by the Fourier transform. This property follows from Bochner's theorem \cite{rudin}. We can see from Eq. (\ref{model}) that the RKS method can also be viewed as a result of convergence of an empirical average towards the expectation value. In fact, Rahimi and Recht have shown, using the Hoeffding inequality \cite{hoeffding}, that the approximated kernel converges uniformly locally to the desired kernel \cite{rahimi-recht0}.

\subsubsection{Quantum enhancement}

According to Wilson \textit{et al.} \cite{wilson}, the RKS method has a performance that is comparable to deep learning methods. The polymer classification results obtained with the QE-RKS model confirm this theory. By using a two-mode linear QPC to compute the cosine kitchen sink \cite{ref42}, we have obtained a simple quantum-enhanced ML  model that has a performance comparable to the classical deep learning model.
The method relies on the circuit shown in the bottom figure of Fig. \ref{fig:LQC1}. The cosine function can easily be obtained by using only one photon with the help of Eq. (\ref{prob}):
\begin{equation}\label{1photon}
\cos x = |\langle 0, 1|{\mathcal{U}(x)}|{1, 0}\rangle|^2 - |\langle 1, 0|{\mathcal{U}(x)}{1, 0}\rangle|^2
\end{equation}

The QE-RKS method can also easily be simulated classically \cite{Chabaud_2021, wilson}.
\section{GKM and QE-RKS results}
    \begin{figure*}[!htpb]
        \centering
        \begin{subfigure}[b]{0.495\textwidth}
            \centering
            \includegraphics[width=\textwidth]{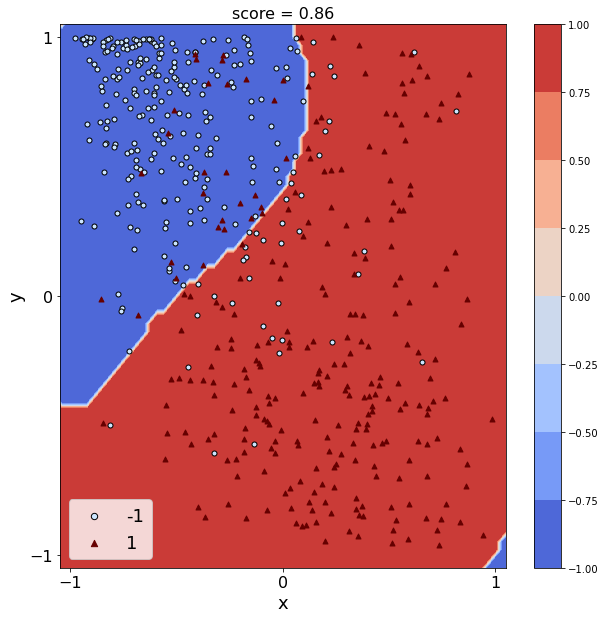}
            \caption[Number of photons: 2]%
            {{\small Number of photons: 2}}    
            \label{fig:}
        \end{subfigure}
        \hfill
        \begin{subfigure}[b]{0.495\textwidth}   
            \centering 
            \includegraphics[width=\textwidth]{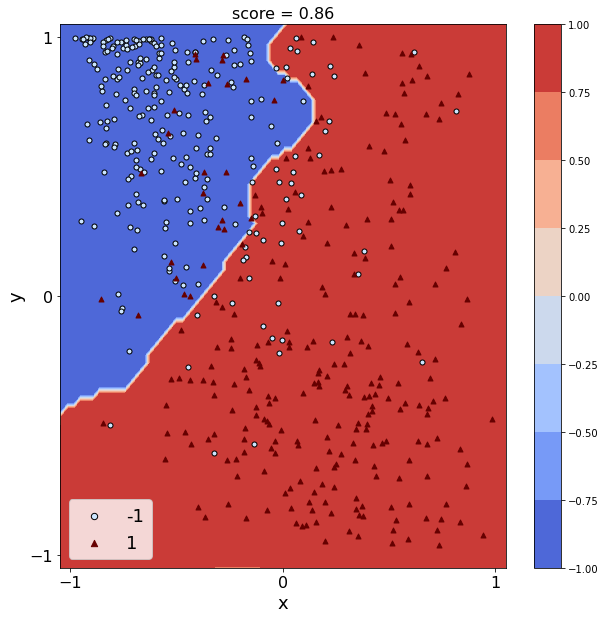}
            \caption[]%
            {{\small Number of photons: 6}}    
            \label{fig:}
        \end{subfigure}
        \hfill
        \begin{subfigure}[b]{0.495\textwidth}   
            \centering 
            \includegraphics[width=\textwidth]{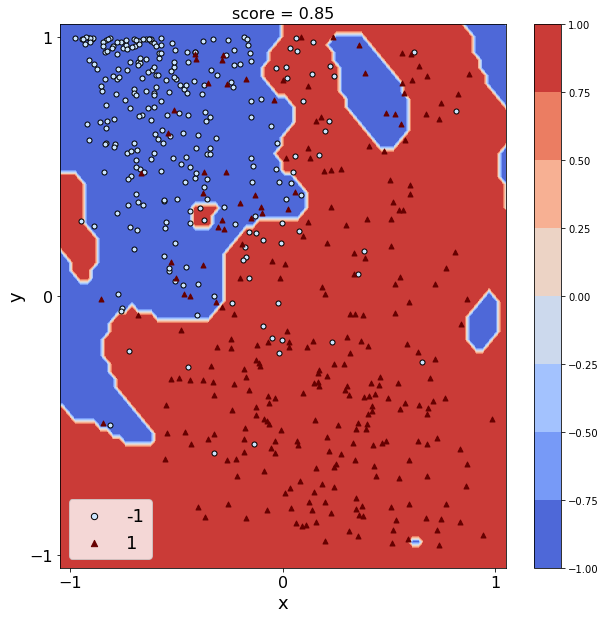}
            \caption[]%
            {{\small Number of photons: 7}}    
            \label{fig:Gaussian}
        \end{subfigure}
         \hfill
         \begin{subfigure}[b]{0.49\textwidth}
            \centering
            \includegraphics[width=\textwidth]{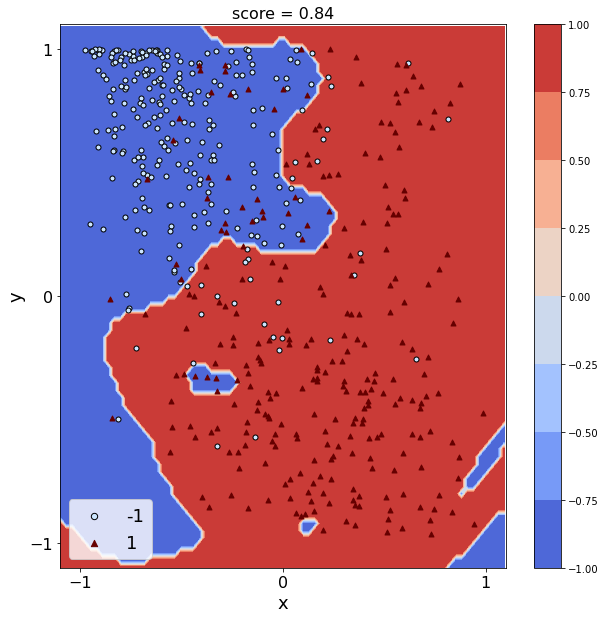}
            \caption[Number of photons: 8]%
            {{\small Number of photons: 8}}    
            \label{fig:mean and std of net14}
        \end{subfigure}
        \vskip\baselineskip
        \caption[ Gaussian Kernel Methods ]
        {\small \small Binary classification of polymers using the trained quantum kernels obtained with the two-mode linear QPC, depicted in the top figure of Fig. \ref{fig:LQC1}, for a different number of input photons $n$ = 2, 6, 7, 8. The regularisation parameter $\alpha$ equals 3.0, 2.5, 2.0, and 4.0, respectively. Test data points are shown in light-blue (VIS [-1] class) and dark-red (NIR [1] class).
        } 
\label{fig:GaussianKernel_method_results_annex_1}
    \end{figure*}
   \begin{figure*}[!htp]
        \centering
        \begin{subfigure}[b]{0.495\textwidth}
            \centering
            \includegraphics[width=\textwidth]{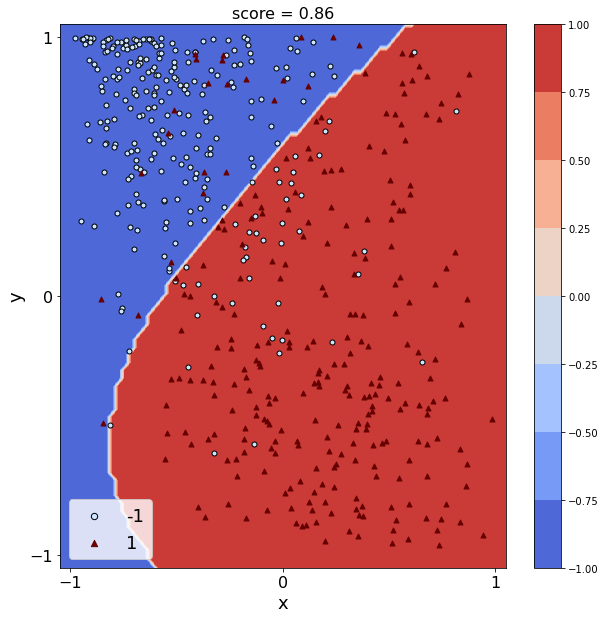}
            \caption[Random Fourier features by  
        sampling from a $\chi$-square distribution.]%
            {{\small  $\chi^2$ distribution, $R$ = 3}} \label{fig:rks_chi}
        \end{subfigure}
        \hfill
        \begin{subfigure}[b]{0.495\textwidth} 
            \centering 
            \includegraphics[width=\textwidth]{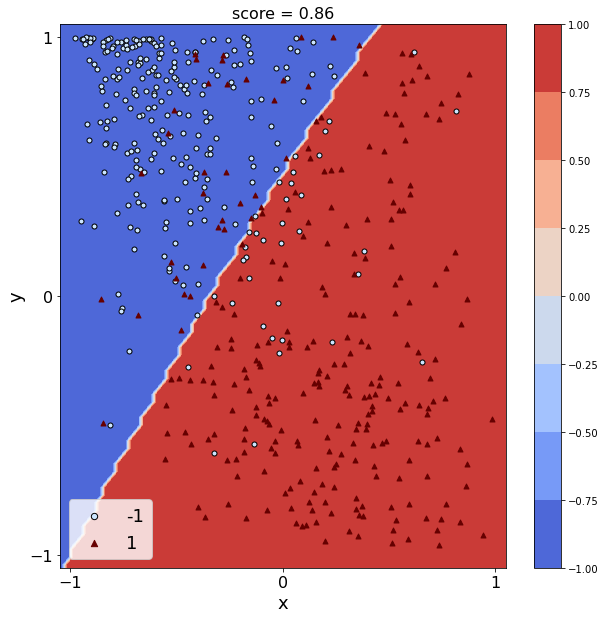}
            \caption[]%
            {{\small  normal distribution, $R$ = 3}}  
            \label{fig:rks_normal}
        \end{subfigure}
                \hfill
        \begin{subfigure}[b]{0.495\textwidth} 
            \centering 
            \includegraphics[width=\textwidth]{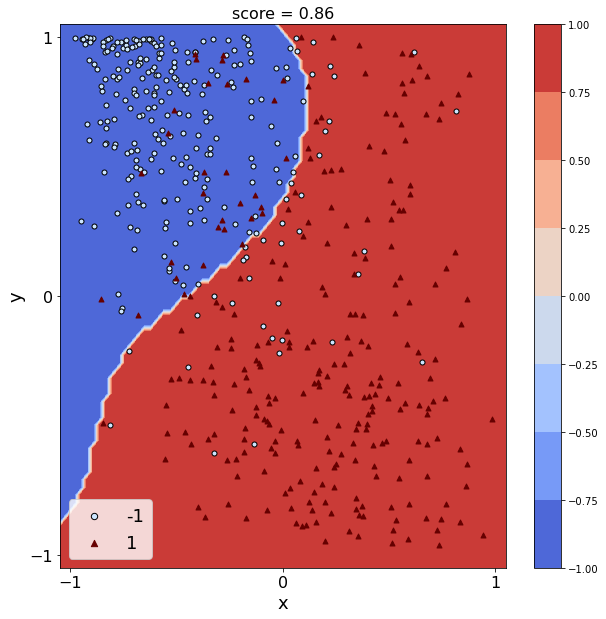}
            \caption[]%
            {{\small  $\chi^2$ distribution, $R$ = 100 }}  
            \label{fig:rks_normal_100}
        \end{subfigure}
                \hfill
        \begin{subfigure}[b]{0.495\textwidth} 
            \centering 
            \includegraphics[width=\textwidth]{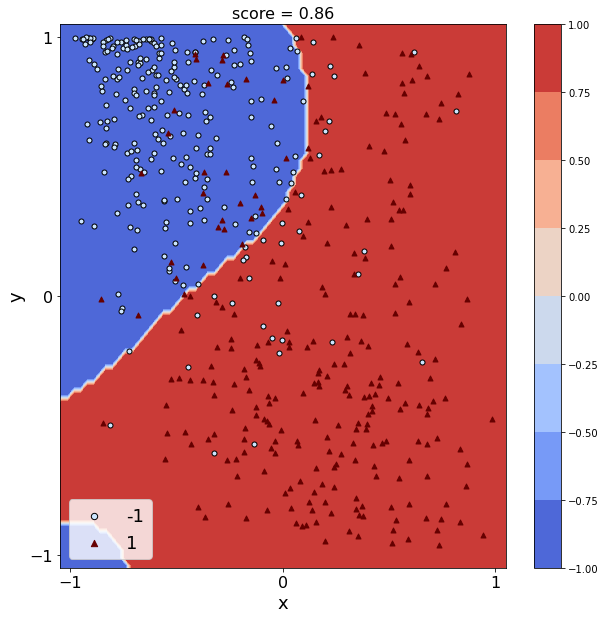}
            \caption[]%
            {{\small normal distribution, $R$ = 100 }} 
            \label{fig:rks_chi_100}
        \end{subfigure}
        \caption[ QE-RKS Methods ]
        {\small Binary classification of polymers using the QE-RKS implemented by the two-mode linear QPC shown in the bottom figure of Fig. \ref{fig:LQC1}. The input Fock state is $|10, 0\rangle$. No regularization weight is applied.
        %The random seed was chosen to be 12345. 
        The random Fourier features are obtained by  
        sampling from either a normal or $\chi^2$ distribution.
        Test data points are shown in light-blue circles (class VIS [-1]) and dark-red triangles (class NIR [1]). 
        }
\label{fig:GaussianKernel_method_results_RKS_10photons}
    \end{figure*}   
   \begin{figure*}[!htpb]
        \centering
    \begin{subfigure}[b]{0.495\textwidth} 
            \centering 
\includegraphics[width=\textwidth]{Figures/RKS_normal_R_3.png}
            \caption[]%
            {{\small $R$ = 3}}  
            \label{fig:rks_r_3}
        \end{subfigure}
        \hfill
        \begin{subfigure}[b]{0.495\textwidth} 
            \centering 
            \includegraphics[width=\textwidth]{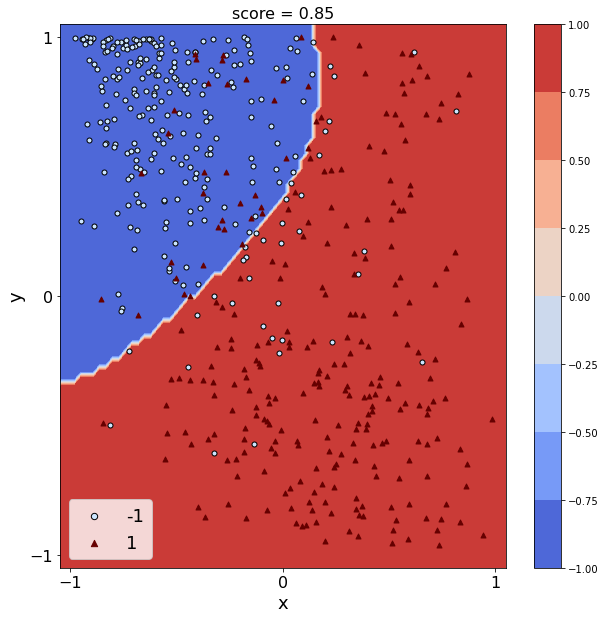}
            \caption[]%
            {{\small $R$ = 10 }} \label{fig:rks_r_10}
        \end{subfigure}
        \hfill
        \begin{subfigure}[b]{0.495\textwidth} 
            \centering 
            \includegraphics[width=\textwidth]{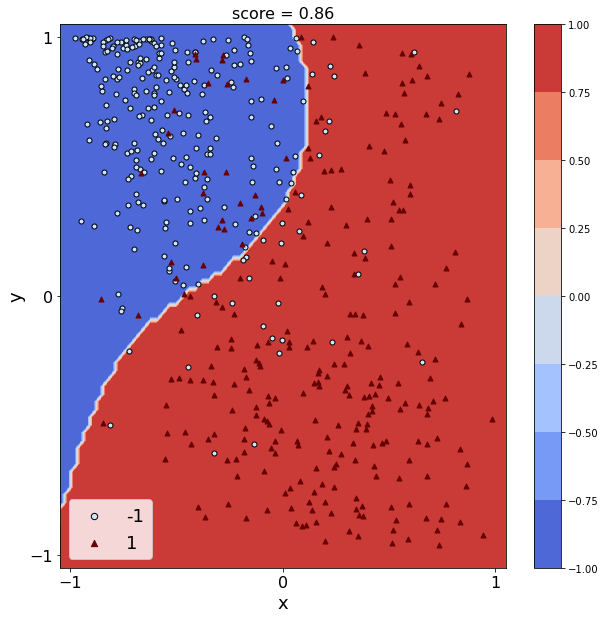}
            \caption[]%
            {{\small $R$ = 20 }}  
            \label{fig:rks_r_10}
        \end{subfigure}
        \hfill
        \begin{subfigure}[b]{0.495\textwidth} 
            \centering 
            \includegraphics[width=\textwidth]{Figures/RKS_normal_R_100.png}
            \caption[]%
            {{\small $R$ = 100 }}  
            \label{fig:rks_r_100}
        \end{subfigure}
        \caption[ QE-RKS Methods 2 ]
        {\small Binary classification of polymers using the QE-RKS implemented by the two-mode linear QPC shown in the bottom figure of Fig. \ref{fig:LQC1}. The input Fock state is $|10, 0\rangle$. No regularization weight is applied. The random Fourier features are obtained by  
        sampling from a normal distribution.
        % The random seed was chosen to be 12345. 
        Test data points are shown in light-blue circles (class VIS [-1]) and dark-red triangles (class NIR [1]). 
        }
\label{fig:GaussianKernel_method_results_RKS_2_10photons}
    \end{figure*} 

\end{document}